\documentclass[preprint2]{aastex62}
\usepackage{comment}

\newcommand{\fluxunit}{erg s$^{-1}$ cm$^{-2}$}
\newcommand{\lumunit}{erg s$^{-1}$}
\newcommand{\fluxhard}{$f_{X,\,2-7\,\mathrm{keV}}$}

\newcommand{\lumhardint}{$L_{X,\mathrm{int},\,2-10 \,\mathrm{keV}}$}
\newcommand{\lumgen}{$L_{X,\,2-10 \,\mathrm{keV}}$}
\newcommand{\lumprimehardint}{$L^{'}_{X,\mathrm{int},\,2-10 \,\mathrm{keV}}$}

\usepackage{enumitem}
\graphicspath{{./}{figures/}}

\received{\today}
\revised{AAAA}
\accepted{AAAA}
\submitjournal{ApJ}

\shorttitle{Obscured AGN in Disguise as Low-Luminosity AGN}
\shortauthors{Lambrides et al.}

\begin{document}

\title{A Large Population of Obscured AGN in Disguise as Low Luminosity AGN in \textit{Chandra} Deep Field South}

\author[0000-0002-0786-7307]{Erini L. Lambrides}
\affil{Department of Physics \& Astronomy, Johns Hopkins University,\\Bloomberg Center, 3400 N. Charles St., Baltimore, MD 21218, USA}
\nocollaboration

\author{Marco Chiaberge}
\affil{Space Telescope Science Institute, 3700 San Martin Drive\\
Baltimore, MD 21218, USA}
\affil{Department of Physics \& Astronomy, Johns Hopkins University,\\Bloomberg Center, 3400 N. Charles St., Baltimore, MD 21218, USA}
\nocollaboration

\author{Timothy Heckman}
\affil{Department of Physics \& Astronomy, Johns Hopkins University,\\Bloomberg Center, 3400 N. Charles St., Baltimore, MD 21218, USA}
\nocollaboration

\author{Roberto Gilli}
\affil{INAF$-$ Osservatorio di Astrofisica e Scienza dello Spazio di Bologna, Via P. Gobetti 93/3, 40129 Bologna, Italy
\\
Bologna, Italy}
\nocollaboration

\author{Fabio Vito}
\affil{Instituto de Astrofísica and Centro de Astroingeniería, Facultad de Física, Pontificia Universidad Católica de Chile, Casilla 306, Santiago 22, Chile}
\affil{Chinese Academy of Sciences South America Center for Astronomy, National Astronomical Observatories, CAS, Beijing 100012, China}
\nocollaboration

\author{Colin Norman}
\affil{Space Telescope Science Institute, 3700 San Martin Drive\\
Baltimore, MD 21218, USA}
\affil{Department of Physics \& Astronomy, Johns Hopkins University,\\Bloomberg Center, 3400 N. Charles St., Baltimore, MD 21218, USA}
\nocollaboration



\begin{abstract}

Population synthesis models of actively accreting super-massive black holes (or active galactic nuclei -- AGN) predict a large fraction that must grow behind dense, obscuring screens of gas and dust. Deep X-ray surveys are thought to have provided the most complete and unbiased samples of AGN, but there is strong observational evidence that a portion of the population of obscured AGN is being missed. In this paper we use a sample of AGN derived from the deepest X-ray survey to date, the \textit{Chandra} 7Ms GOODS-South Survey, to investigate the nature of low flux X-ray sources. We make full use of the extensive multi-wavelength coverage of the GOODS-South field, and cross-match our objects with wavelengths from the Radio to the IR. We find the low X-ray flux AGN in our sample have X-ray luminosities that indicate low-luminosity AGN classification, while their radio, infrared and optical counterparts indicate moderate to powerful AGN classification. We find the predicted column densities is on average an order of magnitude higher than the calculated column densities via X-ray detections for X-ray faint sources. We interpret our results as evidence of obscured AGN disguising as low-luminosity AGN via their X-ray luminosities. When we compare the estimation of the obscured AGN space density with and without these objects, we find a difference of 40\% in the lowest X-ray luminosity regime probed by our sample.

\end{abstract}

\keywords{Active Galaxies -- X-rays -- Optical -- Infrared -- Radio -- Sky Survey -- Obscured AGN}

\section{Introduction} \label{sec:intro}

Theoretical models of galaxy formation predict that massive galaxies should have high star formation rates and larger gas reservoirs than that which is observed. It has been postulated that actively accreting supermassive black holes (SMBHs), known as active galactic nuclei (AGN), can inject energy into the gas and expel it and/or prevent it from cooling and collapsing into stars through a mechanism called feedback \citep[e.g.][]{bower06,croton,heckmanaa}. The ubiquity of SMBHs in the center of galaxies and the large energy release per gram of matter accreted onto the SMBH makes AGN feedback the most promising star formation regulation mechanism. Furthermore, star-formation and SMBH growth have similar evolutionary tracks \citep[see][for a review]{madau}. Theory suggests that feedback from growing SMBHs/AGN is able to successfully reproduce the properties of local massive galaxies \citep[see][for review]{silk12}, and explain the observed galaxy scaling relations and the quenching of star-formation in massive galaxies \citep[e.g.][]{silk98,fabian99,king03,hopkins06,illustris2017}. 

Some models of galaxy evolution and AGN feedback explain the observed scaling relations between SMBHs and galaxy host properties via a merging scenario. In these scenarios, AGN are triggered due to the gravitational torques produced as a result of the merger funnelling gas to the central parsecs of the galaxy. A key component of these models is the majority of SMBH growth is occurring behind large column densities, $N_\mathrm{H} > 10^{23}$ cm$^{-2}$ \citep[e.g.][]{cattaneo05,hopkins08, blecha2018}. These obscured sources are inherently difficult to observe, but their relative contribution to the total number of AGN can be estimated via AGN synthesis models for the cosmic X-ray background \citep[e.g.][]{comastri95,gilli2001,treister05,gilli07,akylas12,ananna19}. Directly observing obscured AGN is possible, but emission at wavelengths less than $2$ $\micron$ are significantly attenuated by the obscuring material. Over a wide range of energies (i.e 0.2--200~keV), X-ray observations are thought to provide one of the most reliable methods of selecting AGN and estimating the amount of obscuration \citep[e.g.][]{brandt05,xue11,liu}; however this is not always true, as \citet{comastri11} and \citet{donley12} show that even some of the deepest X-ray surveys miss a substantial fraction of heavily obscured objects.

Obscured AGN can also be identified in the mid-infrared (MIR) due to the reprocessing of the obscured UV emission \citep[e.g.][]{lacy04,houck05,weedman06,yan07,polletta08,stern12,yan13,mateos13}. As noted in \citet{hickreview}, color-color diagnostics may provide high completeness but only modest reliability due to sources not always having a prominent AGN component. Thus AGN hosted in strongly star-forming galaxies may not be identified. This limitation is compounded by the fact that at high-redshifts ($z > 2$), star formation and AGN activity peak. Aside from AGN identification, disentangling obscured vs un-obscured AGN from MIR colors alone is challenging due to the similarity between these two classes of AGN in their MIR SEDs \citep[e.g.][]{buch06,mateos12,asmus14,hickox17}; thus, the combination of large and deep MIR and X-ray surveys are needed to build a large, statistically robust sample of obscured AGN. 

The deepest X-ray survey to date is the \textit{Chandra} Deep Field South (CDFS) survey which was centered on the GOODS-S region.  Due to the severe amount of Compton scattering and absorption which attenuates the X-ray emission at the lower X-ray energies probed by \textit{Chandra}, data at other other wavelengths must be used to quantify the level of potential AGN obscuration. The obscuring medium which absorbs the X-ray continuum photons re-radiates the energy at MIR wavelengths. The combination of X-ray and IR data has been critical in estimating the amount of obscuration in X-ray surveys with energies $< 10$~keV \citep[e.g.][]{daddi07,donley08,fiore09}. Many studies using X-ray selected AGN select AGN as objects with measured luminosities of $L_{X} > 10^{42}$~\lumunit\ to avoid contamination from galaxies for which the X-ray luminosity is dominated by star formation. To fully understand the AGN population, it is essential to properly account for the possibility that sources with low observed X-ray luminosity may in fact be moderately to heavily obscured. In this paper we investigate the nature of these low luminosity sources. 

In \autoref{sec:sample} we describe the data acquisition and sample properties. In \autoref{sec:results} we present comparisons between the X-ray, radio, IR, and optical counterparts. In \autoref{sec:disc} we discuss the implications of the existence of these sources in two different examples, and we summarize our findings in \autoref{sec:conclusion}. We use an $h = 0.7$, $\Omega_{m} = 0.3$, $\Omega_{\Lambda} = 0.7$ cosmology throughout this paper. We use the k-sample Anderson-Darling mid-rank statistic to test the null hypothesis that two samples are drawn from the same population, and report the test statistic (D$_{ADK}$) significance level at which the null hypothesis for the provided samples can be rejected. 

\section{Sample Selection} \label{sec:sample}

\begin{figure}
\centering
\includegraphics[]{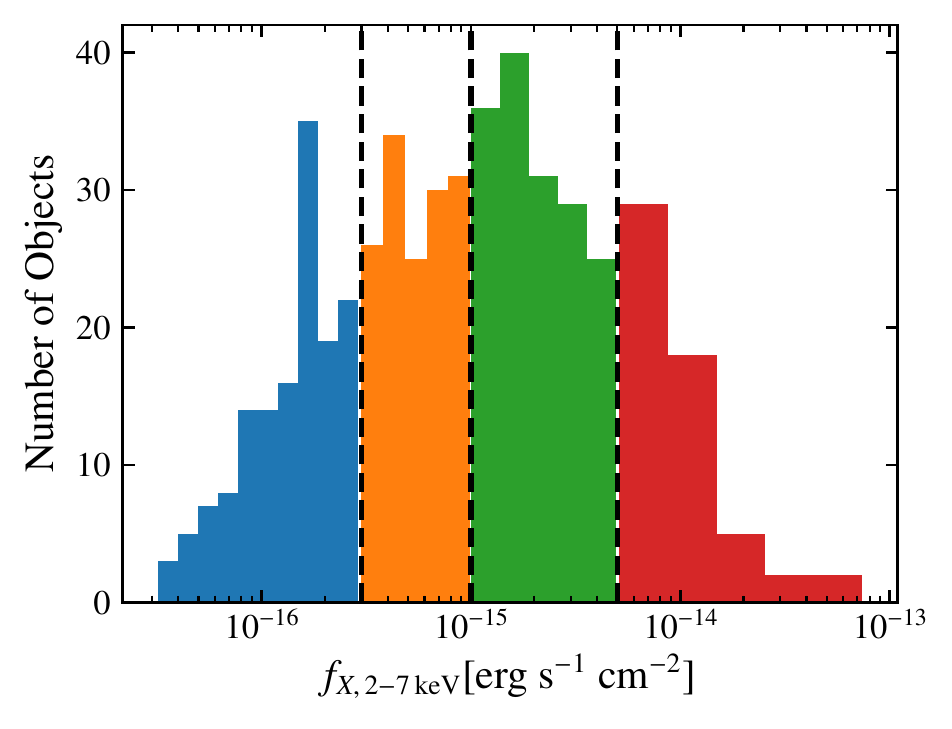}
	\caption{Hard X-ray Flux Distribution of Our Sample: The distribution is split into four bins. The color definitions remain consistent through out this paper.}
\label{fig:bindef}
\end{figure}

\begin{figure}
\centering
\includegraphics[scale=.6]{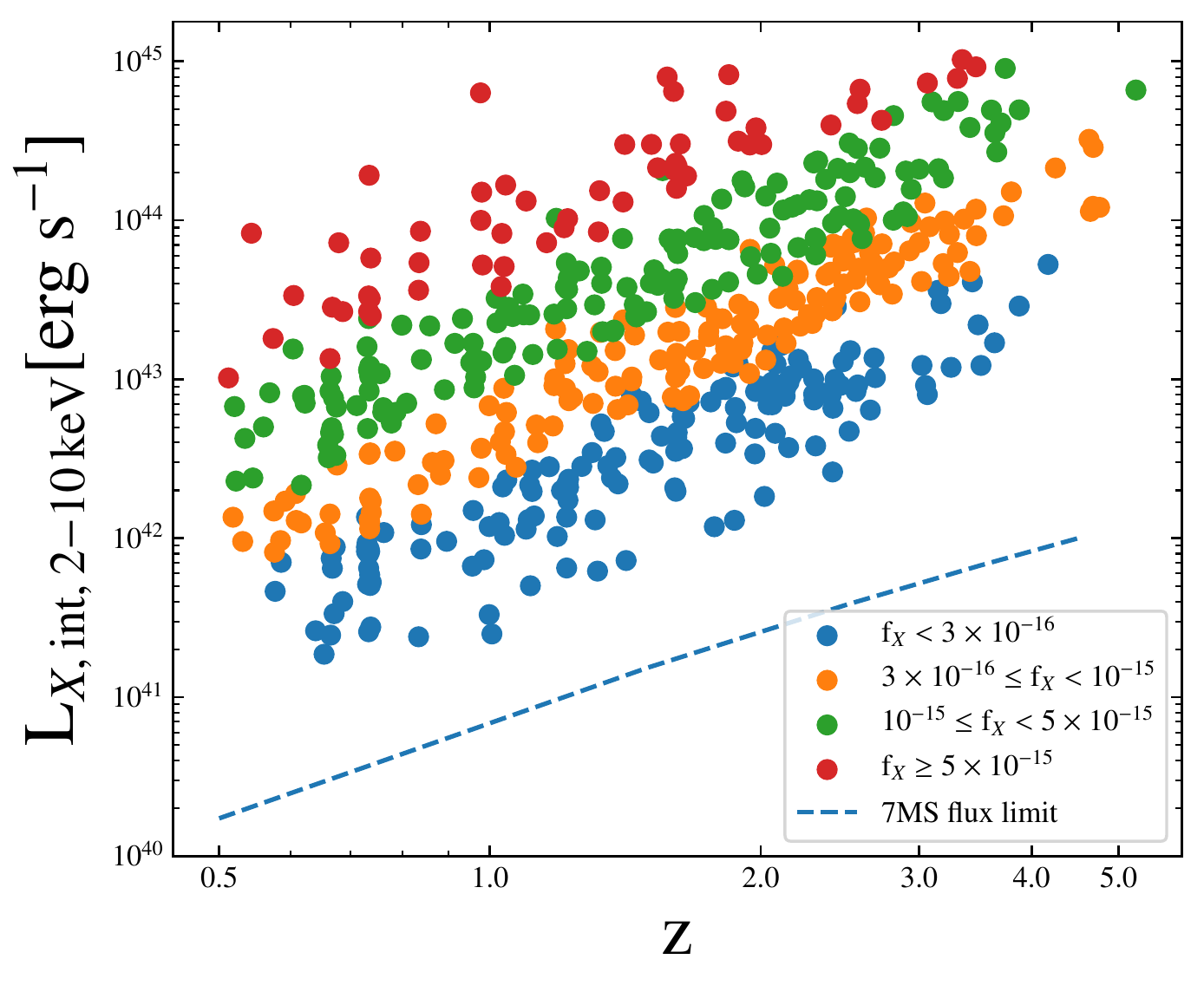}
	\caption{Redshift versus Absorption Corrected or Intrinsic X-ray Luminosity. The median and mean redshifts for our sample are 1.6 and 1.7 respectively. The points are color-coded by their flux bin. The blue dashed line corresponds to the mean \textit{Chandra} 7Ms flux limit \citep{luo17}.}
\label{fig:z_v_lxc}
\end{figure}

The sample discussed in this paper is derived from the deepest X-ray survey to date, the 7Ms exposure \textit{Chandra} Deep Field South (CDFS) survey which covers a total area of 484.2~arcmin$^{2}$ \citep[hereinafter L17]{luo17}. The 7Ms CDFS catalogue contains 1008 sources analyzed in three energy regimes: 0.5--7.0 keV (full), 0.5--2.0 keV (soft), and 2--7 keV (hard). We select 523 CDFS sources that have redshifts  $> 0.5$, were detected in both the full band and hard band, and are labeled as AGN in the L17 catalog. We use the criterion of z $> 0.5$ to maximize the selection of objects in an epoch where we expect the greatest evolutionary effects. In L17, the sources are classified as AGN if they fulfill one of the photometric and/or spectroscopic criteria below:

\begin{enumerate}[label=(\alph*)]
    \item A source with intrinsic luminosity L$_{0.5-7.0 \mathrm{keV}}$ $\geq 3 \times 10^{42}$ ergs s$^{-1}$.
    \item A source with $\Gamma \leq$ 1.0, where $\Gamma$ is the effective photon index and a value of $\leq$ 1.0 is indicative of an obscured AGN.
    \item A source with an X-ray to optical flux ratio of log($f_{\mathrm{X}}/f_{\mathrm{R}}$) $>$ -1 where the X-ray flux is the FB and the R flux is provided in L17.
    \item A source with a factor of 3 or more X-ray emission over the level expected from pure star-formation as traced by the rest radio 1.4 GHz luminosity \citep{alexander05}.
    \item A source with broad emission and/or high-excitation emission lines in the optical spectrum via the cross-matched spectroscopically identified AGN catalogue in \citet{szokoly04}, \citet{mignoli05},\citet{silverman10}.
    \item A source with an X-ray to NIR flux ratio log($f_{\mathrm{X}}/f_{\mathrm{K}_{s}}$) $>$ -1.2.
    
\end{enumerate}{}

As noted in \citet{xue11} and L17, the above criteria are effective but not complete in identifying AGNs. In particular, these selection methods may not capture the lowest luminosity or most obscured AGN. Thus, there may be a fraction of sources classified as "Galaxies" in L17, which in actuality host an AGN. For the scope of this paper, we present our results on the galaxies classified as AGN only, but explore the objects classified as "Galaxies" in the Discussion.

We split the sample into four measured hard X-ray flux bins:
\begin{enumerate}
    \item \fluxhard\ $ < 3 \times 10^{-16}$ \fluxunit
    \item $3 \times 10^{-16}$ \fluxunit\ $ \leq $ \fluxhard\ $ < 1 \times 10^{-15}$ \fluxunit
    \item $1 \times 10^{-15}$ \fluxunit\ $ \leq $ \fluxhard\ $ < 5 \times 10^{-15}$ \fluxunit
    \item \fluxhard\ $ \geq 5 \times 10^{-15}$ \fluxunit. 
\end{enumerate}

The values of the first three bins were chosen to contain roughly equal numbers of objects and correspond to flux limits of other X-ray surveys in this field. The highest flux bin contains roughly 50 objects. The value of 3 $\times 10^{-16}$ \fluxunit, $1 \times 10^{-15}$ \fluxunit corresponds to the flux limit at 50\% sky coverage for the \textit{Chandra} 4Ms 2-8 keV \citet{xue11}, XMM 3.45 Ms GOODS-S 2-10 keV \citet{xmm4ms} surveys respectively. We show the distribution of the hard X-ray flux values as presented in L17 and the bin definitions used throughout this paper in \autoref{fig:bindef}. 

Since the main focus of this paper is to understand the nature of the low-flux sources in the context of AGN classification, we choose the highest energy X-ray band available because it should be least affected by obscuration, and thus, a more accurate indicator of the intrinsic AGN power. Furthermore, we apply an estimate of the intrinsic absorption as derived by L17 to the hard X-ray flux values. The hard X-ray band fluxes as presented in L17 are not absorption corrected, but L17 provide an estimate of the intrinsic absorption which they apply to their full band luminosities, as follows. 

The X-ray spectrum of an AGN can be described by a power law: the photon number density takes the form $N(E) \sim E^{-\Gamma}$ where $\Gamma$ is the photon index and $E$ the photon energy. In L17, they estimate the power-law photon index $\Gamma_\mathrm{eff}$ from the hard to soft band ratios, where $\Gamma_\mathrm{eff}$ includes Galactic absorption. L17 then uses the Portable, Interactive, Multi-Mission Simulator (\texttt{PIMMS}, \citealt{pimms}) to estimate the intrinsic absorption. By assuming that the intrinsic power law spectrum has a fixed photon index of 1.8 modified by Galactic absorption, any value smaller is likely caused by intrinsic absorption ($N_\mathrm{H}$). We then use the estimated $N_\mathrm{H}$ tabulated in L17 to derive the intrinsic hard X-ray luminosity. Finally, we use \texttt{PIMMS} to re-calculate the intrinsic hard X-ray luminosities over the rest 2--10 keV energy band which we define as \lumhardint. We modify the energy band so that our work may be directly comparable to similar studies. In \autoref{fig:z_v_lxc}, we show the \lumhardint\ values as a function of redshift. The points are color-coded by their flux bin. The blue dashed line corresponds to the mean \textit{Chandra} 7Ms flux limit derived in L17, 3.6$\times 10^{-17}$\fluxunit, re-calculated over the 2-10 keV energy range. In the following sections we describe the collection of the multi-wavelength data we use in this work, with the aim of investigating the nature of the low-flux X-ray AGN sources.       

\begin{figure}
\centering
\includegraphics[scale=.35]{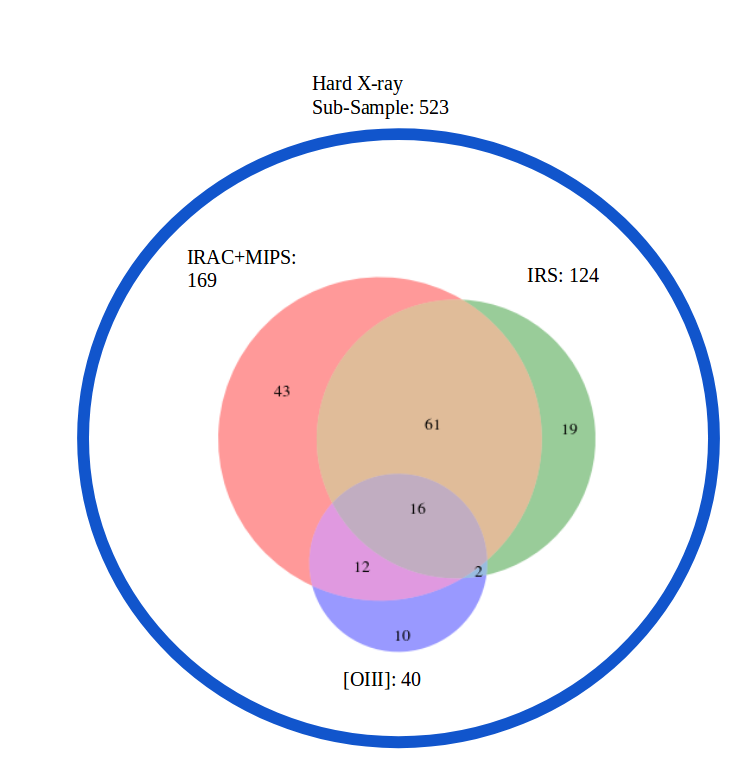}
	\caption{Cross-Match Summary: After selecting all the sources from \citet{luo17} that have ${z > 0.5}$, full and hard band detections and are classified as AGN via their catalogue (as represented by large black circle), we then present the summary of the cross-matching statistics of these 486 objects to the IR and optical data used in \autoref{sec:results}.} 
\label{fig:xmatch}
\end{figure}

\begin{figure*}
\centering
\includegraphics[scale=1]{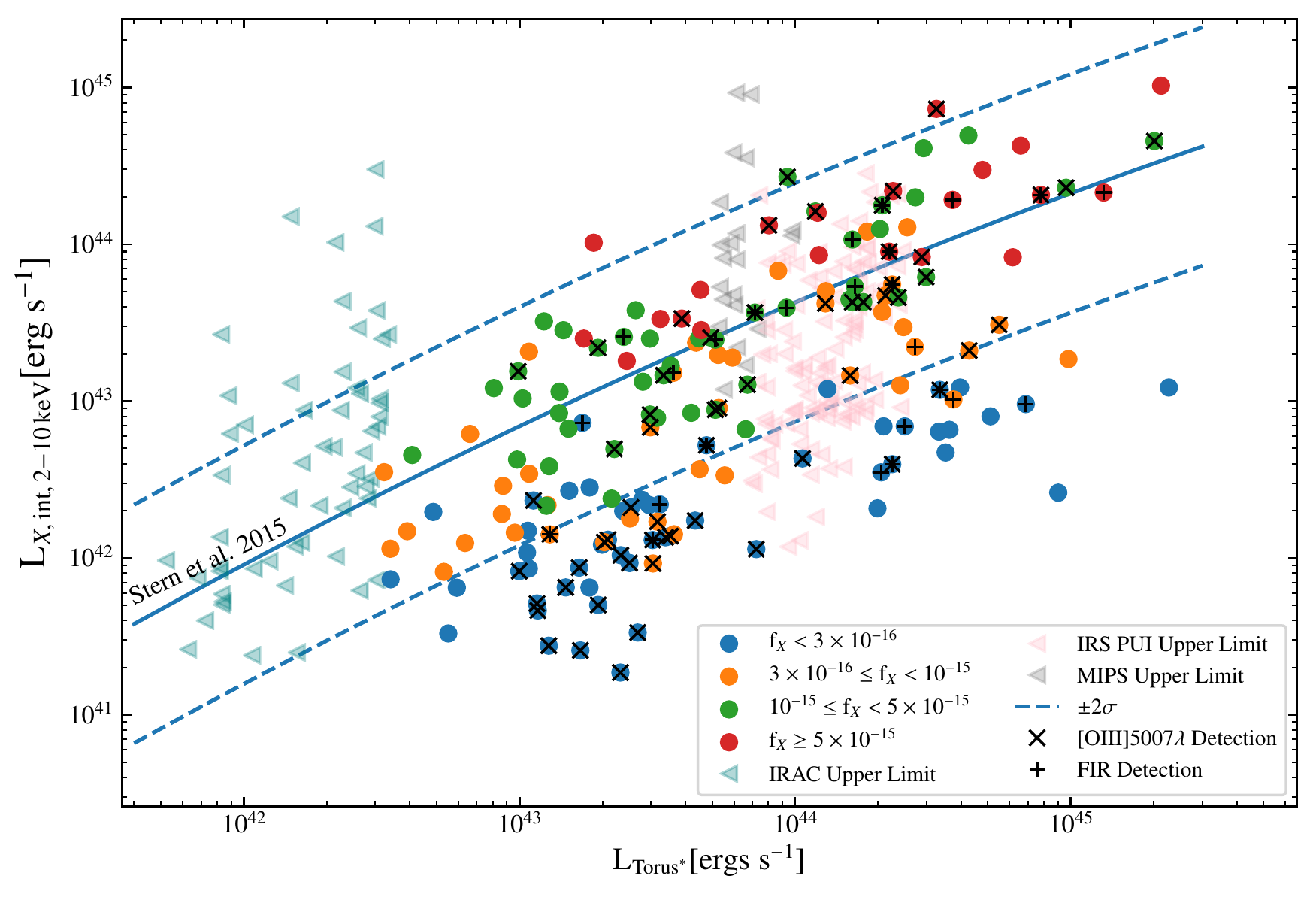}
	\caption{L$_{\mathrm{Torus}^{*}}$ versus \lumhardint: We calculate the rest frame L$_{\mathrm{Torus}^{*}}$ values by splitting the sample into bins of redshift, and using the observed IR flux that corresponds to a rest-frame flux in between 3.4~$\micron$ and 5.7~$\micron$  The points are color-coded by X-ray flux. The blue solid line is the luminosity dependent relationship from S15. The dashed blue lines are the 2$\sigma$ dispersion from the \citet{marchesi2016} sample. The gray open faced triangles are the MIR upper-limits. Objects with } 
\label{fig:lmir_v_lx}
\end{figure*}

\subsection{Infrared Measurements}

We cross-match our X-ray sample to IR catalogues to quantify the effect of varying levels of obscuration on our X-ray fluxes. We use \textit{Spitzer} Infrared Array Camera (IRAC) (3.4~\micron, 4.5~\micron, 5.8~\micron, 8.0~\micron) data, the \textit{Spitzer} peak-up imager (PUI) on the Infra-red Spectrograph (IRS) instrument \citep{houck05} 16 \micron\ data, Multi-band Imaging Photometer for \textit{Spitzer} (MIPS) (24~\micron) data, \textit{Herschel} Photodetector Array Camera and Spectrometer (PACS) (100~\micron, 160~\micron) data. The \textit{Spitzer} IRAC and \textit{Herschel} data were taken from the GOODS-\textit{Herschel} survey catalogue, where \textit{Herschel} flux densities and uncertainties were obtained from point source fitting using \textit{Spitzer} 24~\micron\ detections positions as a prior \citep{elbaz11}. L17 provide the optical counterparts to the CANDELS + 3D\textit{HST} combined catalogue \citep{skelton14}. We cross-match our sources to the \citet{elbaz11} catalogue, which also provided associated GOODS counterpart coordinates, using the optical counterpart coordinates with a 1 arcsecond search tolerance. The \textit{Spitzer} IRS PUI 16~\micron\ detections were also found using the optical counterpart coordinates with a 1 arcsecond cross-match search tolerance to the 16~\micron\ GOODS-S catalogue \citep{irspui}. We find 169 X-ray (32.3\%) source matches in all four IRAC bands and MIPS 24~\micron, 124 (24\%) matches in \textit{Spitzer} IRS PUI 16~\micron\ data, and 76 (14.5\%), 62 (11.9 \%) objects have PACS 100~\micron\ and 160~\micron\ detections respectively.We note the majority of the analysis in \autoref{sec:results} is constrained to X-ray sources with IRAC bands and MIPS 24~\micron\ detections.Within this X-ray, MIR sub-sample, over 50\% of objects have both Herschel detections.

\subsection{Optical Measurements}

Another probe of AGN power is the strength of high ionization optical lines. L17 provide the counterparts to the CANDELS + 3D\textit{HST} combined catalogue \citep{skelton14}. We use these coordinates from the L17 catalogue to perform a cross-match with a 0.2 arcsecond tolerance to the rest-frame color catalogue \citep{skelton14}, emission line catalogue \citep{momcheva16}, and ACS/WFC3IR images \citep{skelton14}. We have overlap with 167 objects with $> 5\sigma$ detections with ACS F435W photometry, and 253 objects with $> 5\sigma$ detections with WFC3-IR F160W photometry. In the emission line catalogue, we have 40 objects with $> 2\sigma$ [\textsc{OIII}] detections.

\subsection{Radio Measurements}

Radio emission is present in both AGN and star-forming dominated galaxies. The L17 catalogue provides 1.4 GHz fluxes via the Very Large Array (VLA) survey centered on the CDFS field. We find 94 objects above the 5$\sigma$ flux density limit of 37~$\mu$Jy. For the detected objects, we use the redshifts provided in L17 and calculate the rest frame luminosity for each object assuming a radio spectral index of $\alpha = 0.8$ where $\alpha$ is defined as $f_{\nu} \propto \nu^{-\alpha}$. For the un-detected objects we calculate the upper limit using the limiting flux of the GOODS-S VLA survey \citep{vla}, from which the upper limit fluxes are derived. In \autoref{fig:xmatch}, we show a summary of all the cross-matching results of our L17 sub-sample with the other wavelengths.

\section{Results} \label{sec:results}

\subsection{X-ray and Rest-Frame 5 \micron\ Continuum} \label{sec:MIR}

As stated in \autoref{sec:intro}, the combination of hard X-ray data and mid-infrared (MIR) data offers one of the best probes of obscuration in AGN host galaxies. X-ray emission is one of the most unambiguous signatures of AGN activity, and in obscured AGN, the material that attenuates the X-ray emission is expected to emit in the MIR. A measurement of a bright, un-obscured source in both the X-ray and MIR allows for empirical relationships to be derived between these quantities. The MIR contains features which can be attributed to AGN and/or SF processes. Between 3.2 \micron\ to 5.7 \micron, AGN torus emission dominates over MIR SF processes \citep{nenkova,kirkpatrick,lambrides}. Previous studies have used SED decomposition or templates to calculate the rest-frame emission in this region \citep{delmoro12, stern15}. Due to the uncertainties introduced with these methods, we instead take advantage of our large sample and its multi-wavelength properties. 

\begin{figure*}
\centering
\includegraphics[]{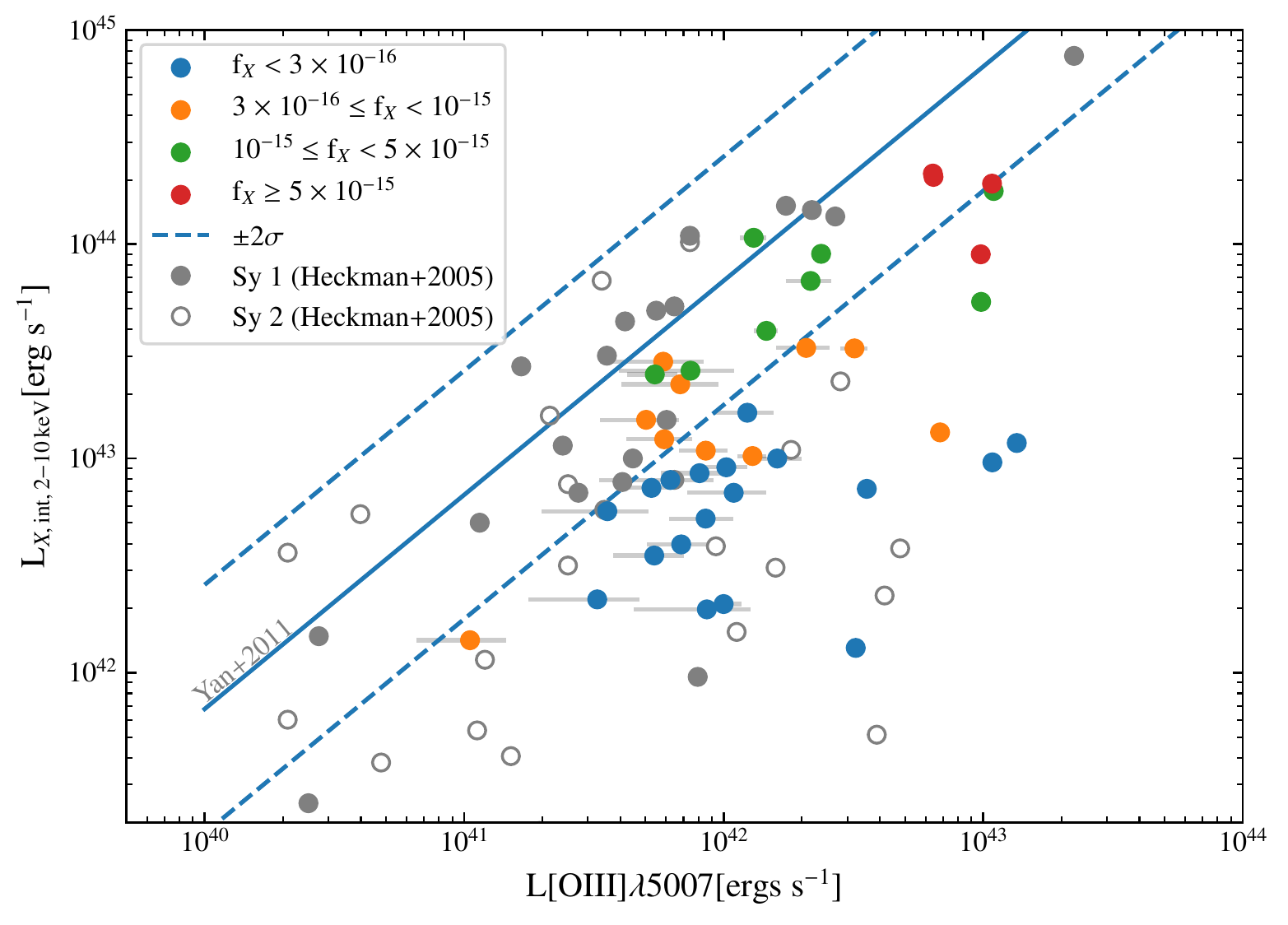}
	\caption{\lumhardint vs L[\textsc{Oiii}]$\lambda 5007$: The points are color-coded by \fluxhard. The blue solid line is the relationship parametrized for Type 1 AGN in the AEGIS sample \citep{yan11}. The error bars are the L[\textsc{Oiii}]$\lambda$5007 1$\sigma$ confidence intervals via \citet{momcheva16}. The filled, and open faced gray circles are type 1, type 2 AGN respectively from \citet{heckman2005}.} 
\label{fig:loiii_v_lxflux}
\end{figure*}

We infer the rest-frame 5 \micron\ continuum region luminosities to directly measure the emission in this spectral region. For regions of redshift where  $0.5 < z \leq 1.5$, $1.8 < z \leq 3.1$, and $z \geq 3.1$, we use the observed IRAC 8 \micron, IRS PUI 16 \micron, and MIPS 24 \micron\ luminosities respectively. This corresponds to rest-frame luminosities in the 3.2 \micron\ to 5.7 \micron\ continuum region, depending on the object's redshift, and refer to these luminosities as L$_{\mathrm{Torus}^{*}}$ for simplicity. We use the nomenclature Torus$^{*}$ because although we are not estimating the entirety of the IR Torus luminosity, we expect the AGN torus luminosity to dominate over SF processes in this wavelength regime. 

We test the prediction of the AGN torus luminosity dominating the flux emission in the wavelengths used to calculate the L$_{\mathrm{Torus}^{*}}$ values on a local AGN and Starburst sample. \citet{lambrides} uniformly analyzed all AGN and SF galaxy ever observed with the \textit{Spitzer} Infrared Spectrograph \citep{irs}. As similar studies have shown, \citet{lambrides} found, for even  low-luminosity (L$_{24 \micron}$ $<$ $10^{42}$ [erg s$^{-1}$]) AGN, the PAH 6.2 $\micron$ equivalent width (EQW) is an excellent indicator of AGN contribution to the MIR: the lower the 6.2 \micron\ EQW, the more the spectrum is dominated by an AGN component. The EQW classifier was able to separate highly star-forming Ultra-Luminous Infrared Galaxies with and without an AGN. Using the spectra and other cross-matched data provided in \citet{lambrides}, we calculate L$_{\mathrm{Torus}^{*}}$ using the same approach as in this work. We find L$_{\mathrm{Torus}^{*}}$ is as good as the 6.2 \micron\ classifier. Performing a Spearman rank correlation on L$_{\mathrm{Torus}^{*}}$ (normalized by K band luminosity to account for mass difference) and PAH 6.2 \micron\ EQW, we measure an anti-correlation (p-value $<$ .001). As an additional check we used the EQW to estimate the L$_{5\micron}$ from the AGN component alone or  L$_{5\micron, \mathrm{AGN}}$ which is is defined as L$_{5\micron}$ $\times$ (1.0 - 0.54/EQW). L$_{5\micron}$ is derived from the spectra and is an integrated quantity between rest-frame 4.8$\micron$ to 5.2$\micron$. The factor of (1.0 - 0.54/EQW) is set to equal 1 when the EQW $>$ 0.54, for this represents 100\% of the emission is being powered by SF \citep[see][for details of EQW to MIR AGN power estimates]{armus07,petric11,stierwalt14}. We find a 0.1 dex agreement between the L$_{5\micron, \mathrm{AGN}}$ values and the L$_{\mathrm{Torus}^{*}}$ values. The only region of the parameter space where L$_{\mathrm{Torus}^{*}}$ may fail is in objects where L$_{\mathrm{Torus}^{*}}$ is completely dominated by SF. However, such objects would cluster around the solid yellow line in \autoref{fig:l50ltor}, which we do not observe. 

\begin{figure*}
\centering
\includegraphics[scale=1 ]{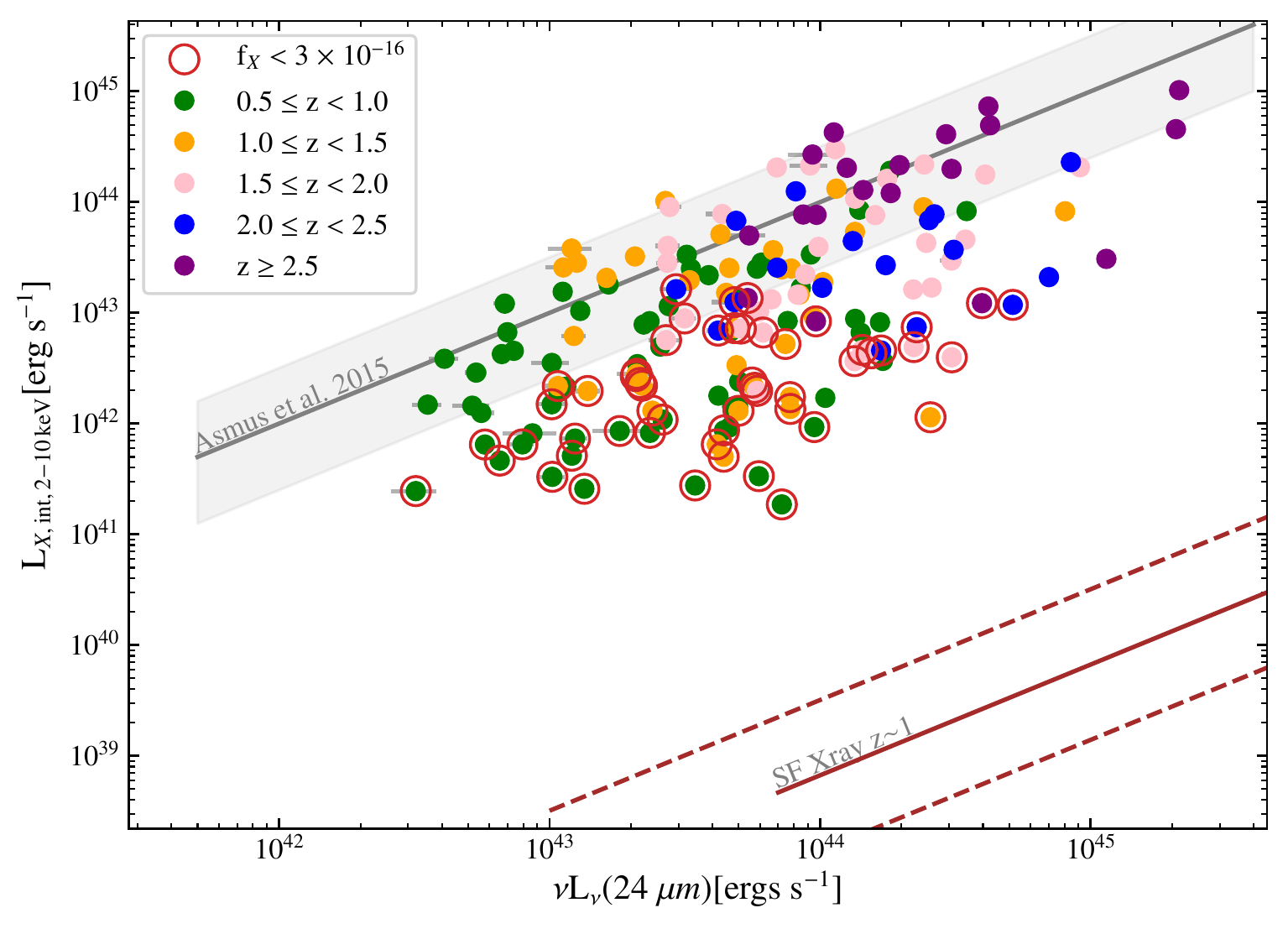}
	\caption{24 \micron\ vs \lumhardint: The points are color-coded by redshift. The gray line surrounded by the grey shaded area is the \citet{asmus15} relation for nuclear 18~\micron\ and intrinsic 2--10 keV luminosities with the dispersion of values from their sample of un-obscured AGN. The red solid  and dashed lines is the X-ray to 24\micron\ relationship for SF galaxies and $\pm 2 \sigma$ respectively for a sample of $z \sim 1$ SF galaxies from \citet{sym14}. This relationship is adapted from \citet{sym14} by converting the L$_{8-1000 \micron}$ values to the 24\micron luminosity using the conversion presented in \citet{calzetti10}. The red-circle indicate objects that are in the lowest X-ray flux bin.} 
\label{fig:l24_v_lxc}
\end{figure*}

We show the relationship between L$_{\mathrm{Torus}^{*}}$ and \lumprimehardint\ in \autoref{fig:lmir_v_lx}. The blue solid line is the luminosity dependent absorption corrected X-ray, $\nu$L$_{\nu}$(6 $\micron$) relationship from \citet{stern15}, hereafter S15, which parametrizes the relationship as  $\log L(2-10 \mathrm{keV}) = 40.981 + 1.024x - 0.047x^{2}$ where $x \equiv \log (\nu$L$_{\nu}$(6 $\micron$)). We chose the S15 relation due to the similar method in which they derived the equivalent L$_{\mathrm{Torus}^{*}}$ luminosity and the large luminosity range their sample covers. We find over 90$\%$ of our objects with the lowest X-ray fluxes, f$_{X,2-7 \mathrm{keV}}$ $< 10^{-16}$ \fluxunit, are $\geq 2 \sigma$ below the S15 relation. We compute the Anderson-Darling statistic to test whether the lowest flux bin is drawn from the same population of \lumprimehardint\ to L$_{\mathrm{Torus}^{*}}$ values from the rest of the flux bins, and we find the null hypothesis can be rejected (D$_{ADK}$ = 43.7, critical value(1\%) = 6.546, significance = 0.001). The teal, pink, and gray triangles are the IRAC, IRS PUI, and MIPS upper-limits respectively. We calculate the upper-limits using the flux limits provided for the relevant MIR wavelength used in the L$_{\mathrm{Torus}^{*}}$ calculation. The flux limit for  observed IRAC 8~\micron, IRS PUI 12~\micron, and MIPS 24~\micron\ is 1.6~$\mu$Jy, 65~$\mu$Jy, and 20~$\mu$Jy respectively. We test the effect of upper-limits by performing a censored regression analysis on each X-ray flux bin, and we find the  L$_{\mathrm{Torus}^{*}}$ and \lumprimehardint\ relationship in each flux bin remains unchanged when upper-limits are taken into account.

The tension between the X-ray and L$_{\mathrm{Torus}^{*}}$ for the lowest X-ray flux objects suggests that i) The low X-ray flux objects are intrinsically weak AGN with a non-AGN component contributing to the MIR luminosity, or ii) The low X-ray flux objects are moderately to heavily obscured AGN. With regards to scenario i), any non-AGN component in these systems would most likely arise from SF processes. In section \ref{sec:optical}, we compare the X-ray emission to a direct probe of AGN power that can be less effected by obscuration as compared to the X-rays: the [\textsc{Oiii}]$\lambda 5007$ luminosity. In section \ref{sec:sfrest} we test whether the excess MIR emission for the lowest flux sources can be attributed to a low-luminosity AGN in a host galaxy with extreme amounts of SF.

\subsection{X-ray and [OIII]$\mathit{\lambda 5007}$ Luminosities} \label{sec:optical}

The luminosity of emission lines formed in the narrow line region, such as [\textsc{Oiii}]$\lambda 5007$, can be used as a quasi-isotropic indicator of AGN power \citep{brinch04,heckman2005,lamassa2010}. [\textsc{Oiii}]$\lambda 5007$ is one of the strongest narrow forbidden lines and is emitting in a region far from the dusty torus. We check if a robust optical line indicator of AGN power is consistent with the X-ray emission of our objects. The [\textsc{Oiii}]$\lambda 5007$ feature may be attenuated due to either nuclear or host galaxy obscuration. Thus, without correction, an observed [\textsc{Oiii}]$\lambda 5007$ luminosity may be thought of as a lower limit. Using the [\textsc{Oiii}] fluxes derived from \textit{HST} grism spectroscopy provided in \citet{momcheva16}, we compare the calculated [\textsc{Oiii}] luminosities to \lumhardint\ in \autoref{fig:loiii_v_lxflux}. In \autoref{fig:loiii_v_lxflux}, we also plot the relationship of L[\textsc{Oiii}]$\lambda 5007$ versus \lumhardint\ for a sample of optically selected type 1 AGN \citep{yan11}. Furthermore, for comparison, we additionally plot the type 1 and type 2 AGN sample from \citet{heckman2005}. The \citet{heckman2005} sample is not corrected for nuclear obscuration. We find our results from \autoref{fig:lmir_v_lx} are consistent with \autoref{fig:loiii_v_lxflux}: 85\% of our lowest flux objects are at least 2$\sigma$ below the \citet{yan11} relation, and in the same parameter space of the \citet{heckman2005} type 2 AGN sample. We compute the Anderson-Darling statistic to test whether the lowest flux bin is drawn from the same population of \lumprimehardint\ to L[\textsc{OIII}]$\lambda 5007$ values from the rest of the flux bins, and we find the null hypothesis can be rejected (D$_{ADK}$ = 8.30, critical value(1\%) = 6.546, significance = 0.001). The inconsistency between the X-ray and the [\textsc{OIII}] emission observed for a substantial fraction of X ray sources strongly hints at them not being truly low-power AGN. In the next section, we follow-up on this hypothesis by checking whether these apparently under-luminous X-ray sources have an extra component in the MIR due to an extremely large amount of star formation.

\subsection{Do the low X-ray flux objects have significant SF?} \label{sec:sfrest}

\begin{figure}
\includegraphics[scale=.9]{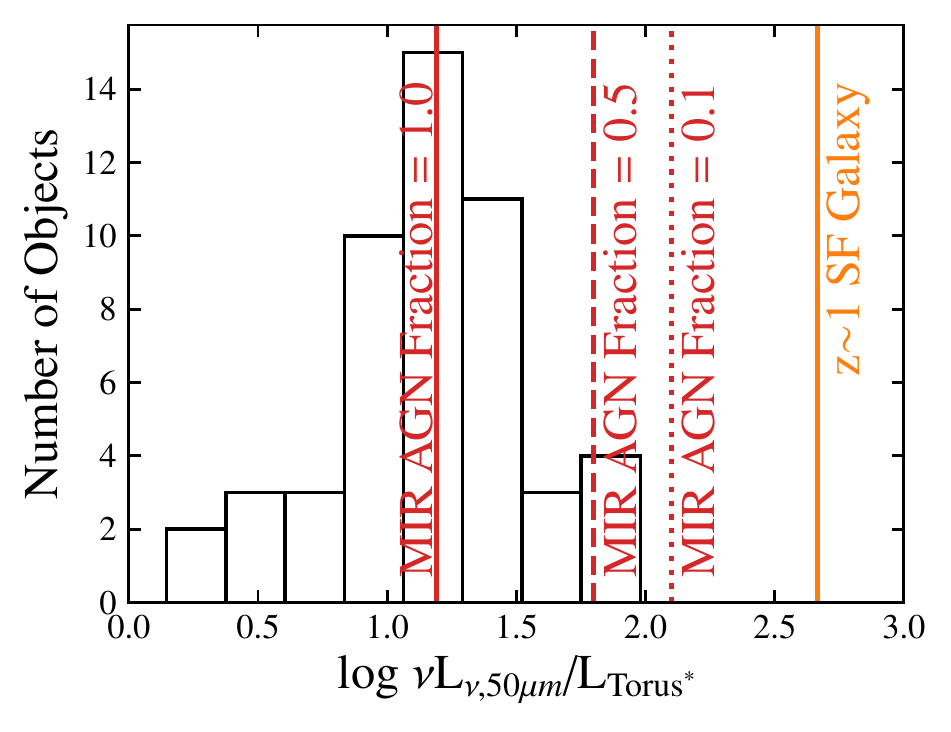}
	\caption{L$_{50\micron}$ to L$_{\mathrm{Torus}^{*}}$: The solid, dashed and dotted red lines are the ratio values for the \citet{kirkpatrick2015} templates with MIR AGN fractions of 1.0, 0.5, 0.1 respectively. The solid orange line is the ratio value for the \citet{kirkpatrick} $z\sim1$ SF galaxy template.} 
\label{fig:l50ltor}
\end{figure}

Monochromatic continuum luminosity at 24~\micron\ is commonly used to trace star-formation due to the warm dust associated with high-mass star-forming regions emitting at this wavelength \citep{calzetti}. On the other hand, SF processes also contribute to the soft and hard X-ray components \citep{persic02}. X-ray emission in SF galaxies is predominately from gas in the ISM heated by stellar winds and supernova and point sources such as X-ray binaries. For star-forming galaxies, very high IR luminosities (L$_{IR}$ $>$ 10$^{46}$ \lumunit) must be observed in order to correspond to L$_{X,\mathrm{2-10\ kev}}$ $>$ 10$^{42}$ \lumunit\ \citep{sym14}. Conversely, for galaxies with an AGN, the 24~\micron\ continuum luminosity may be significantly contaminated with reprocessed light from the central engine. Even more importantly, in AGN the X-ray emission tightly traces the power of the central engine, unless the central engine is obscured. Thus, the relationship between X-rays and the IR will vary significantly between SF and AGN dominated galaxies.

In \autoref{fig:l24_v_lxc}, we show the relationship between \lumhardint\ and the observed 24~\micron\ luminosities for our sample. AGN studies quantifying this relation, or using other MIR continuum measurements, find an almost one to one relationship between these quantities, with minimal scatter ($<$ 1 dex) \citep{gandhi09, asmus15}. The grey shaded region is the range of values for un-obscured AGN adapted from \citet{asmus15}. The red solid  and dashed lines is the X-ray to 24\micron\ relationship for SF galaxies and $\pm 2 \sigma$ respectively for a sample of $z \sim 1$ SF galaxies from \citet{sym14}. This relationship is adapted from \citet{sym14} by converting the L$_{8-1000 \micron}$ values to 24\micron\ using the conversion presented in \citet{calzetti10}. The points are color-coded by redshift, and the points that are circled in red are the lowest-flux bin objects. Similarly to what we showed in the previous sections, we see an apparent inconsistency. A significant fraction of the X-ray sources appear to be under-luminous with respect to their observed 24~\micron\ luminosity. For the redshift range spanned by our sources, the rest frame 24~\micron\ wavelength ranges from 6 to 16~\micron. 

We test if there is a significant dependence between redshift and location of the points with respect to the \citet{asmus15} relation due to our usage of the observed 24~\micron\ fluxes. We quantify the fraction of objects below the \citet{asmus15} relation in each redshift bin, and we find for $0.5 < z < 1.0$: 61\%, $1.0 < z < 1.5$: 56\%, $1.5 < z < 2.0$: 55\%, and $2.0 < z < 2.5$: 71\%, $ z > 2.5$: 21\%. For every redshift bin, excluding the highest bin, the fraction of sources 2$\sigma$ below the \citet{asmus15} relation is between 50\% and 60\%. The lower fraction in the highest redshift bin is most likely due to the difference in sensitivity of the MIPS survey as compared to the 7Ms survey. In fact, as quantified in \citet{elbaz11}, a $ z \sim 3$ galaxy, would need to be at least $1 \times 10^{46}$ \lumunit\ in order to be 5 $\sigma$ above the flux limit of 100~$\mu$Jy in MIPS 24~\micron, and our sample does not contain any such objects.

As seen in \autoref{fig:l24_v_lxc}, the objects that deviate the most from the \citet{asmus15} parameter space are the low-flux X-ray objects, but they are all at least 4$\sigma$ above the SF relation. Furthermore, these very same objects are below the canonical X-ray relations with the L$_{\mathrm{Torus}^{*}}$ and optical line emission (see \autoref{fig:lmir_v_lx}, \autoref{fig:loiii_v_lxflux}).

In addition, we can directly estimate the contribution SF processes may have on the portion of the SED that L$_{\mathrm{Torus}^{*}}$ probes. A common method used in the literature to diagnose the extent that SF processes power the MIR spectrum is via color-selecting diagnostics (e.g \citet{sajina2005,lacy04,stern05,lacy2007,stern12,assef2013}). A potential issue highlighted in the literature is a substantial fraction of X-ray selected AGN being missed in the MIR color-based methods of AGN classification. \citet{donley12} extensively cover the reliability and completeness of both the IRAC color-wedge \citep{lacy04,stern05,lacy2007,stern2012} and their own power-law criteria in the context of the X-ray luminosity of their sources. They find fewer than 20\% of \lumhardint $< 10^{43}$ erg s$^{-1}$ sources are selected as AGN via these methods. They infer the majority of AGN missed by the IRAC wedge and IRAC power-law criteria are lower-luminosity and/or more heavily obscured AGN. Additionally, the IRAC AGN wedge does not reliably select AGNs at higher redshifts. As shown in \citet{kirkpatrick13,kirkpatrick2015}, some z $>$ 0.1 SF galaxies erroneously become selected as AGN as the IRAC bands begin to probe redder wavelengths of the spectrum. In summary, the completeness and reliability issues of color-based methods are exacerbated for the types of objects in our sample: higher-redshift, lower to moderate luminosity AGN.  Due to the large multi-wavelength nature of our sample, we can use rest-frame luminosities and directly compare the regions of the SED that are dominated by SF processes to the region L$_{\mathrm{Torus}^{*}}$ probes.
   
Between 100\micron\ to 160\micron\, the dust spectrum can be approximated by a power law: $f_{\nu} \sim \nu^{\alpha}$. We use the observed 100\micron\ and 160\micron\ to calculate $\alpha$, and extrapolate the 50\micron\ luminosity. The ratio of L$_{50\micron}$ to L$_{\mathrm{Torus}^{*}}$ is smaller in galaxies where AGN dominate the 5\micron\ emission. As shown in \citet{brown}, the peak AGN contamination is in the MIR, and red-wards of 30 \micron, the contribution becomes less significant. 

We calculate the L$_{50\micron}$ to L$_{\mathrm{Torus}^{*}}$ ratios for our sample. In \autoref{fig:l50ltor}, we show the distribution of our values. We also show the ratio for three different MIR AGN templates from \citet{kirkpatrick2015}. An MIR AGN fraction of 1.0 corresponds to a galaxy whose AGN dominates the SED between 5 $\micron$ to $15 \micron$. As the MIR AGN fraction decreases, the SF contribution in this wavelength regime increases. In \autoref{fig:l50ltor}, the solid, dashed and dotted red lines correspond to L$_{50\micron}$ to L$_{\mathrm{Torus}^{*}}$ ratio values for the \citet{kirkpatrick2015} AGN templates respectively with MIR fractions of 1.0, 0.5, and 0.1 respectively. The solid orange line is the ratio value for the z$\sim$1 SF galaxy template from \citet{kirkpatrick13}. 

The median L$_{50\micron}$ to L$_{\mathrm{Torus}^{*}}$ ratio for our sample is 1.1, and is roughly 3 times smaller than the expected SF value. Our results show that L$_{\mathrm{Torus}^{*}}$ is not significantly contaminated with SF emission. This corroborates the idea that our objects do not host intrinsically weak AGN, with a large star-forming component.

\begin{figure}
\includegraphics[scale=.55]{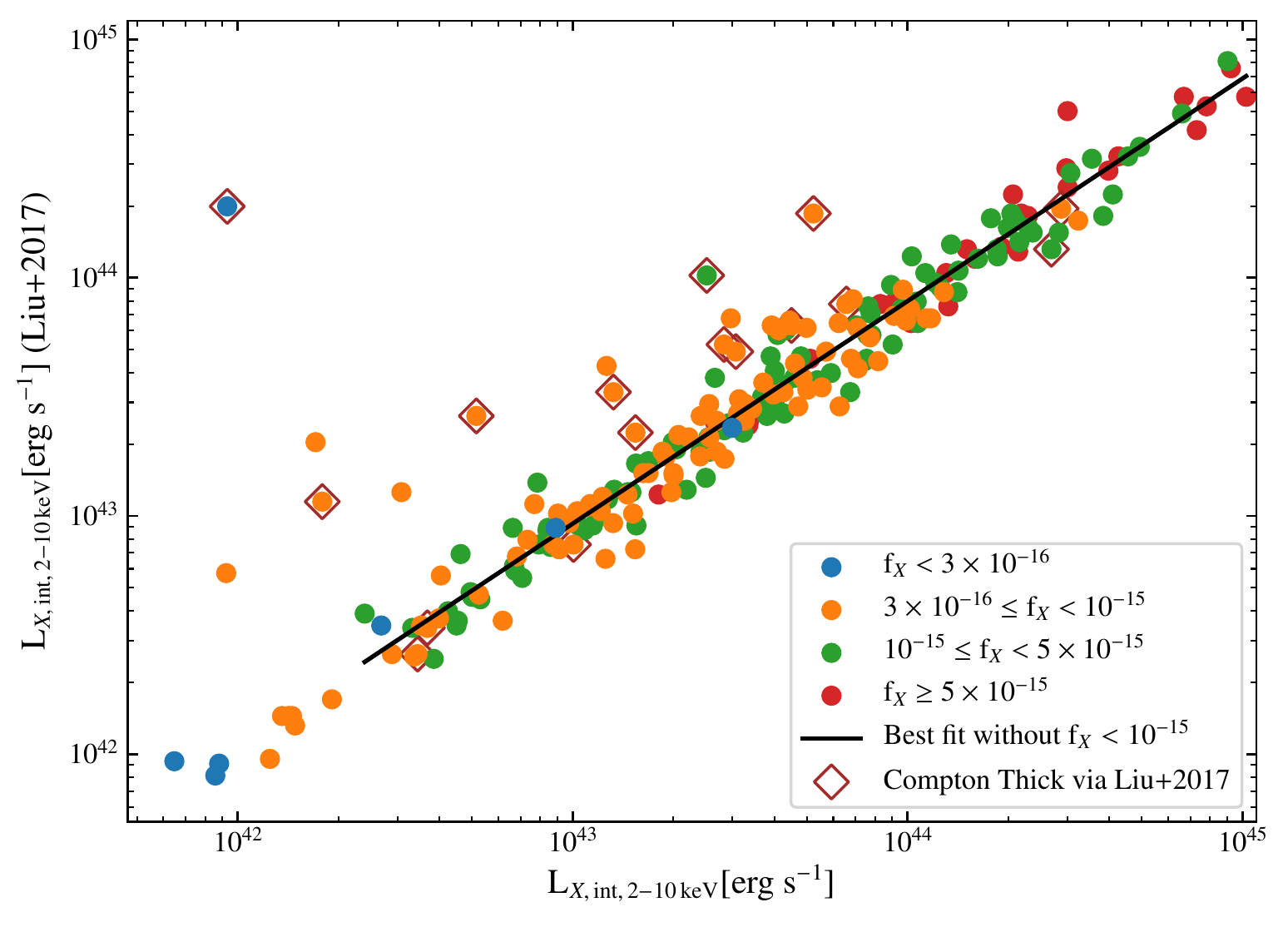}
	\caption{Comparison of our \lumhardint\ and \citet{liu} spectrally derived \lumhardint. The blue colored points are in the lowest X-ray flux bin ($<$ 3 $\times$ 10$^{-16}$ \fluxunit). The black solid line is the best fit relationship for all the objects excluding the two lowest flux bins. The open faced diamonds are the objects spectrally classified as Compton thick AGN via \citet{liu}.} 
\label{fig:liu_luo_diff}
\end{figure} 

\begin{figure*}
\centering
\includegraphics[]{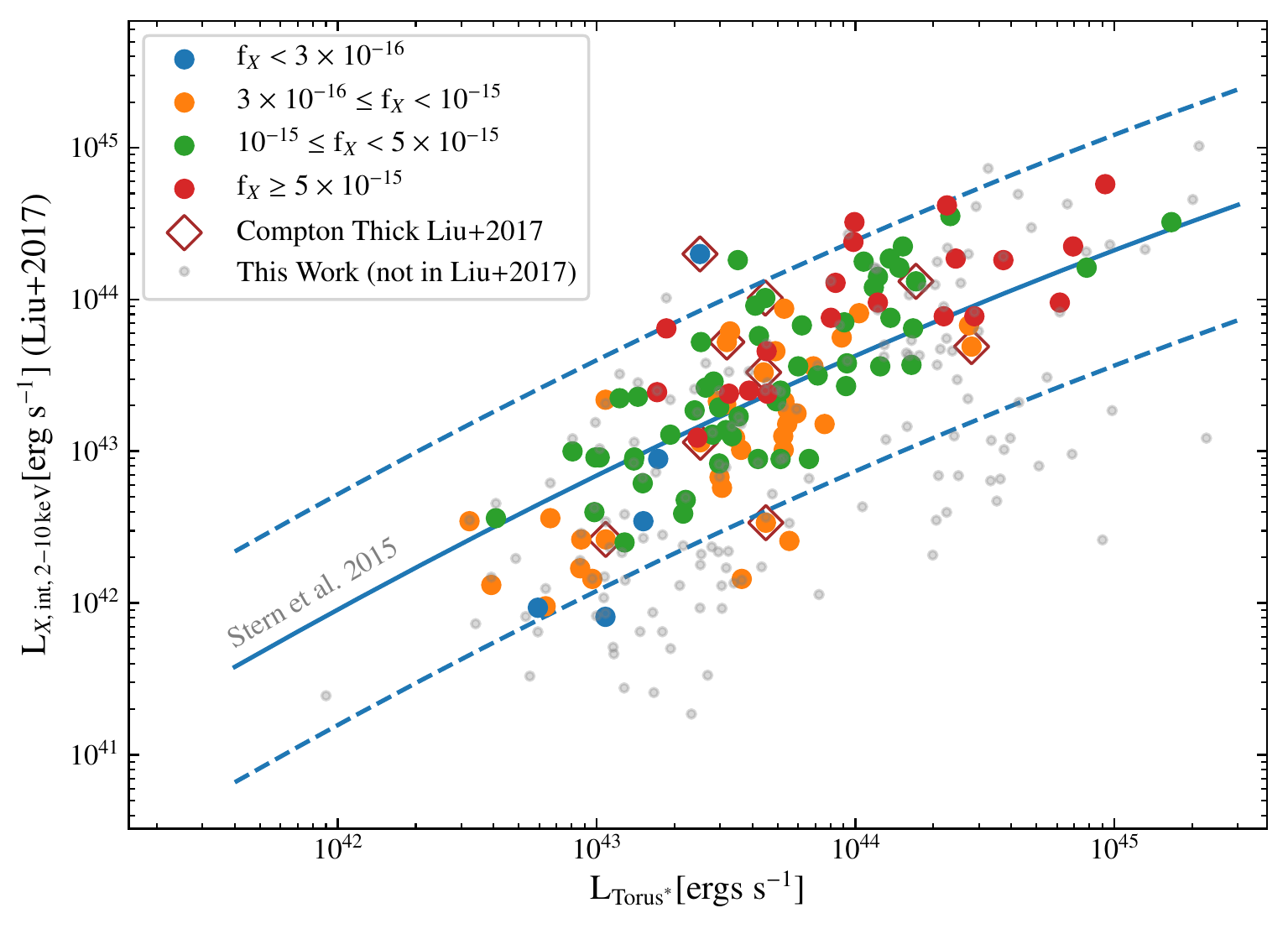}
	\caption{\lumhardint\ vs L$_{\mathrm{Torus}^{*}}$ where the absorption corrected luminosities are provided from \citet{liu}. The points are color-coded by the X-ray flux provided in L17, and the fluxes have units of \fluxunit. The blue solid line is the luminosity dependent relationship from S15. Points surrounded by an open-faced red diamond are classified as Compton thick in \citet{liu}. The grey points are the \lumhardint\ values from \autoref{fig:lmir_v_lx} that did not have enough X-ray counts to be analyzed in \citet{liu}.}
\label{fig:liu_lmir_lxc}
\end{figure*}

\subsection{\textit{Chandra} 7Ms Total Sample versus Spectrally Constrained Sample}

As seen in sections \ref{sec:MIR}, \ref{sec:optical}, and as will be seen in \ref{sec:radio}, the X-ray luminosities derived from simple assumptions are significantly underestimating the intrinsic luminosity of the low flux sources. On the other hand, \citet{liu} performed a detailed spectral analysis on the X-ray bright AGN in the \textit{Chandra} 7Ms sample. Their objects were selected from L17 only if they were classified as AGN and had at least 80 counts in the hard band. This threshold corresponds to a 2--7 keV flux of 2$\times$10$^{-16}$ \fluxunit. They performed a systematic X-ray spectral analysis, with emphasis on constraining intrinsic obscuration. We compare the X-ray properties derived from their 7Ms sub-sample, to the our L17 sub-sample. In \autoref{fig:liu_luo_diff}, we show a comparison of \lumhardint\ of our sample derived from L17 and \lumhardint derived from \citet{liu}. The blue colored points are the lowest X-ray flux bin objects ($< 3  \times$ 10$^{-16}$ \fluxunit). We expect the higher flux bins to be the least affected by the X-ray under-estimation in L17, and thus more consistent with the \citet{liu} analysis. Therefore, the black solid line is the best fit relationship for all the objects excluding the two lowest flux bins. Of the 16 objects classified as Compton thick via \citet{liu} and in our sample, the difference between \lumhardint\ derived in \autoref{sec:sample} and the spectrally derived hard X-ray luminosities is on average $-0.6$ dex. It is important to point out that 44\% of these Compton thick sources are in the lowest X-ray flux bins. We also note that over 78\% of the lowest flux objects in the L17 sample were not spectrally analyzed in \citet{liu} due to their low flux counts.

\begin{figure*}
\centering
\includegraphics[]{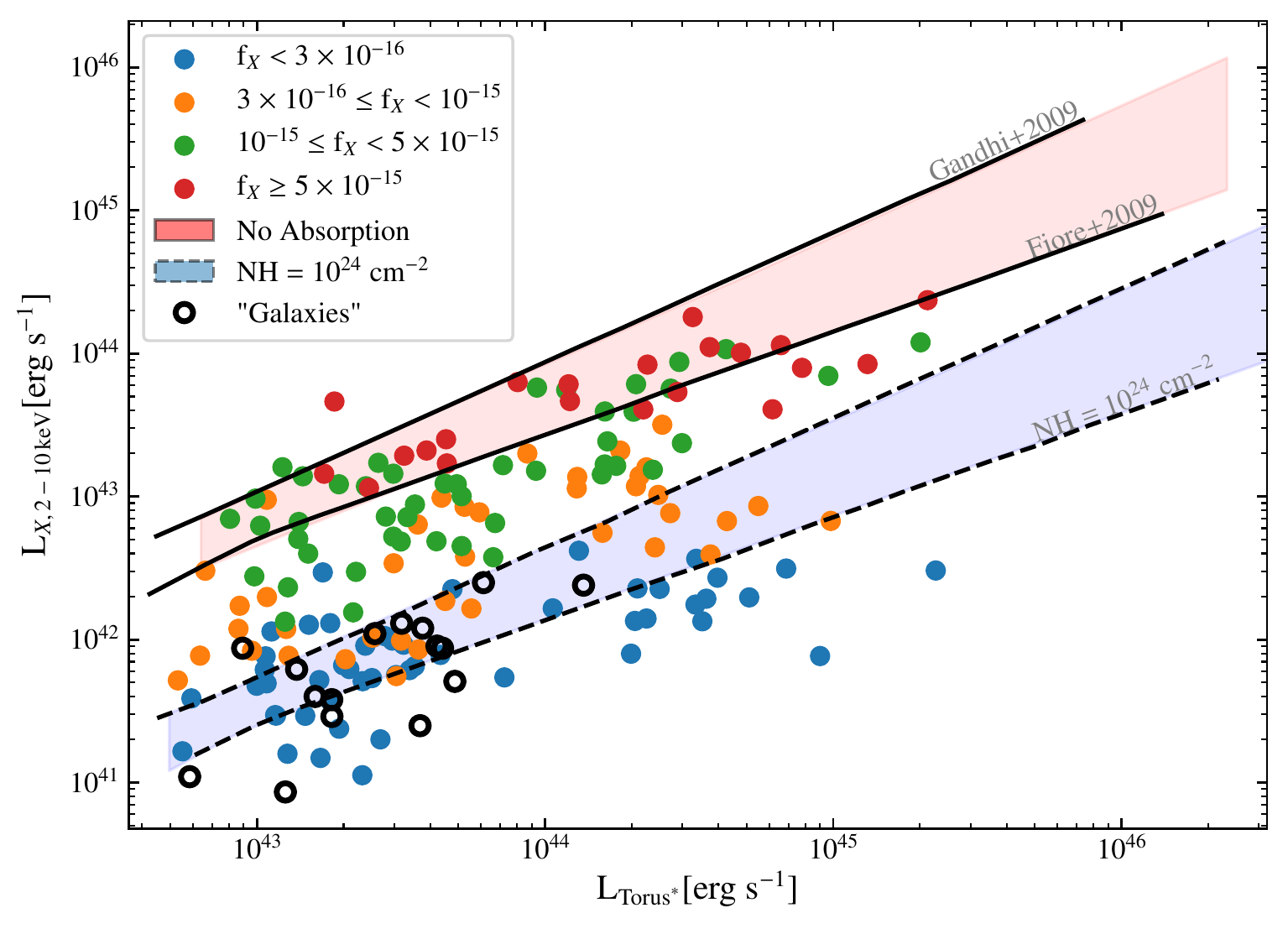}
	\caption{Non-absorption corrected luminosities vs L$_{\mathrm{Torus}^{*}}$: The blue points have the lowest X-ray flux ($<$ 3$\times 10^{-16}$\fluxunit). As adapted by \citet{lansbury15}, the red shaded region indicates the range in intrinsic X-ray, 6 \micron\ AGN luminosity relationships between \citet{gandhi09} and \citet{fiore09}. The blue shaded region indicates the same relationships but where the X-ray luminosity is absorbed by a column density of $N_\mathrm {H} > 10^{24}$ cm$^{-2}$ \citep{lansbury15}. The open black circles are L17 classified "Galaxies" with z $>$ 0.5 and with a detection in the HB.} 
\label{fig:nh_lx}
\end{figure*}

We then compare the \citet{liu} intrinsic hard band X-ray luminosities with L$_{\mathrm{Torus}^{*}}$ in \autoref{fig:liu_lmir_lxc}. The points are color-coded by the X-ray flux provided in L17. The blue solid line is the luminosity dependent relationship from S15. Points surrounded by an open-faced red diamond are classified as Compton thick in \citet{liu}. The grey points are the \lumhardint\ values from \autoref{fig:lmir_v_lx} that did not have enough X-ray counts to be analyzed in \citet{liu}. The \citet{liu} absorption corrected luminosities bring these objects closer or to within 2$\sigma$ of the S15 relationship. Thus, when a more sophisticated X-ray analysis is available the intrinsic absorption estimation yields more accurate luminosities for sources with enough photon counts. The sources in our sample which have the greatest under-estimation of X-ray luminosity have insufficient X-ray counts to perform the above spectral analysis. Thus, when X-ray photon statistics are poor, X-ray vs multi-band diagnostics are necessary to approximate obscuration.

\section{Discussion} \label{sec:disc}

\subsection{The Nature of Low X-ray Flux Sources}

In \autoref{sec:results}, we find a population of low X-ray flux objects whose physical nature is unclear when taking into account the properties of MIR and optical emission. A classification of these sources based on their X-ray luminosity identifies these objects as low-luminosity AGN. However, when only considering the MIR and optical line emission, the same objects are classified as moderate to high luminosity AGN. 
More quantitatively, in \autoref{sec:MIR}, we find 44\% are at least 2$\sigma$ below the expected S15 relationship. Of these objects, 90\% are in the lowest X-ray flux bin. In \autoref{sec:optical}, we find 85\% of our sample have [\textsc{OIII}]$\lambda 5007$ luminosities that are $\geq 2\sigma$ below their predicted \lumhardint values via the \citet{yan11} relationship. In \autoref{sec:sfrest}, we show that the tension between the X-ray luminosities and L$_{\mathrm{Torus}^{*}}$ cannot be explained by an unaccounted for SF component. Thus, we find strong evidence for a large population of obscured AGN disguising as low-luminosity AGN.

The multi-wavelength analysis of this work indicates that over 40\% of our sample has under-estimated intrinsic obscuration. We note that the lowest flux objects correspond to a mean X-ray luminosity of 2.8 $\times 10^{42}$ [ergs s$^{-1}$]. Although in L17 there are multiple criteria that are used to differentiate an X-ray source as an AGN versus an SF galaxy, only one of the seven criteria need to be satisfied for a source to be determined as an AGN. The majority of the criteria, as noted in \autoref{sec:sample}, only capture moderate to high power AGN with the exception of the criterion that the Full band X-ray luminosity is greater than 10$^{42}$ [ergs s$^{-1}$]. High power AGN are rare in the small volume that the CDFS field probes, and thus the most common criteria the X-ray sources satisfy in L17 is the X-ray luminosity threshold. Due to our results indicating that a large fraction of sources may have X-ray luminosities underestimated by at least an order of magnitude, the objects in L17 that are classified as galaxies may also be obscured AGN.

\begin{figure}
\centering
\includegraphics[scale=.75]{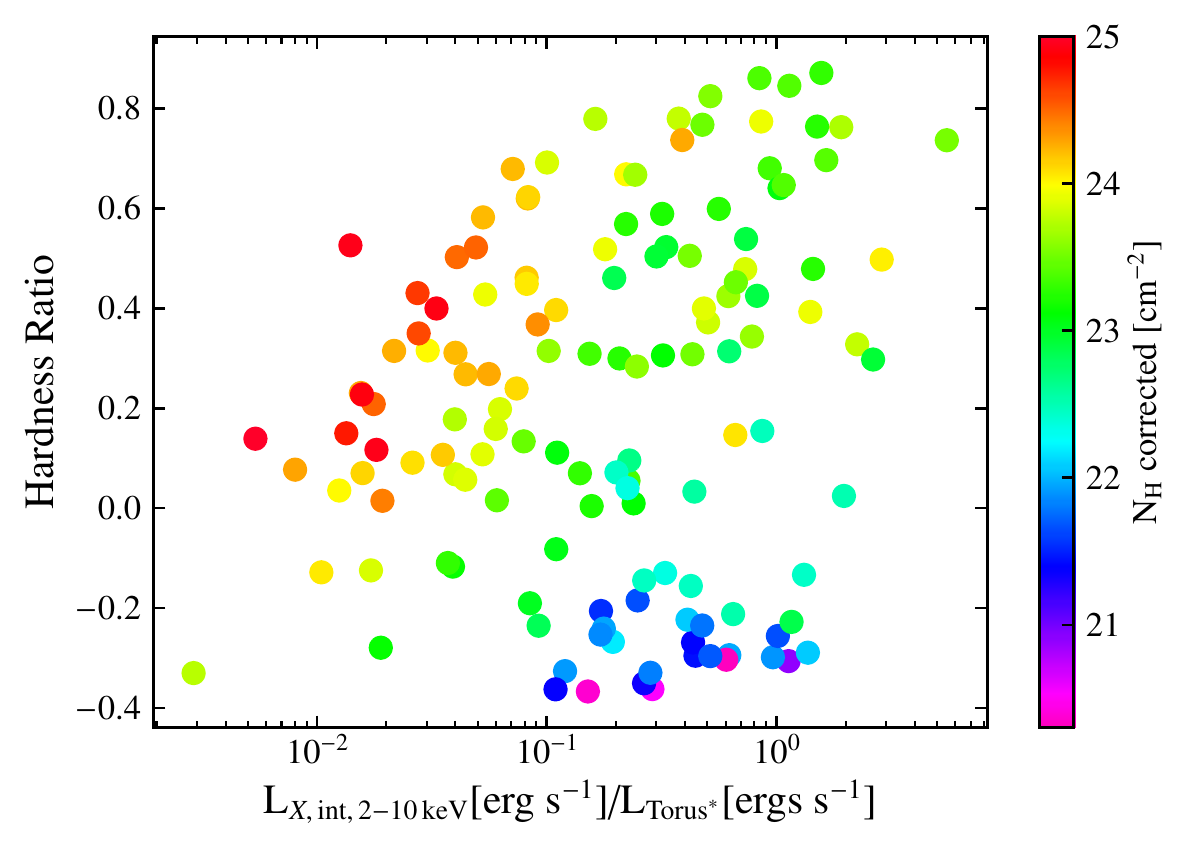}
	\caption{HR as a function of \lumhardint / L$_{\mathrm{Torus}^{*}}$: The points are color coded by our corrected $N_\mathrm{H}$.} 
\label{fig:HR_NH}
\end{figure}

In the entirety of the L17 catalogue, 307 sources are classified as “Galaxies”. Of these 307 sources, we select objects with z $> 0.5$ and detections in the HB in order to be consistent with the L17 classified AGN sub-sample. We use these objects in the analysis moving forward, and label them as “Galaxies”. The “Galaxies” sub-sample consists of 28 sources, where 80\% have a calculated rest-frame L$_{\mathrm{Torus}^{*}}$ values, 20\% have uncorrected [OIII]5007$\lambda$ luminosities greater than 10$^{42}$ [erg s$^{-1}$], and 14\% have VLA 1.4 GHz detections.  Of the “Galaxies” L$_{\mathrm{Torus}^{*}}$ sub-sample, 62\% have both Herschel PACS detections, and a mean, median 50 $\micron$ to 5 $\micron$ luminosity ratio of 1.26, 1.29 respectively. 

In the following sub-sections, we estimate the potentially unaccounted for obscuration and highlight some implications that might arise when one uses the most recent literature X-ray values for these objects. 

\begin{figure*}
\centering
\includegraphics[scale=.9]{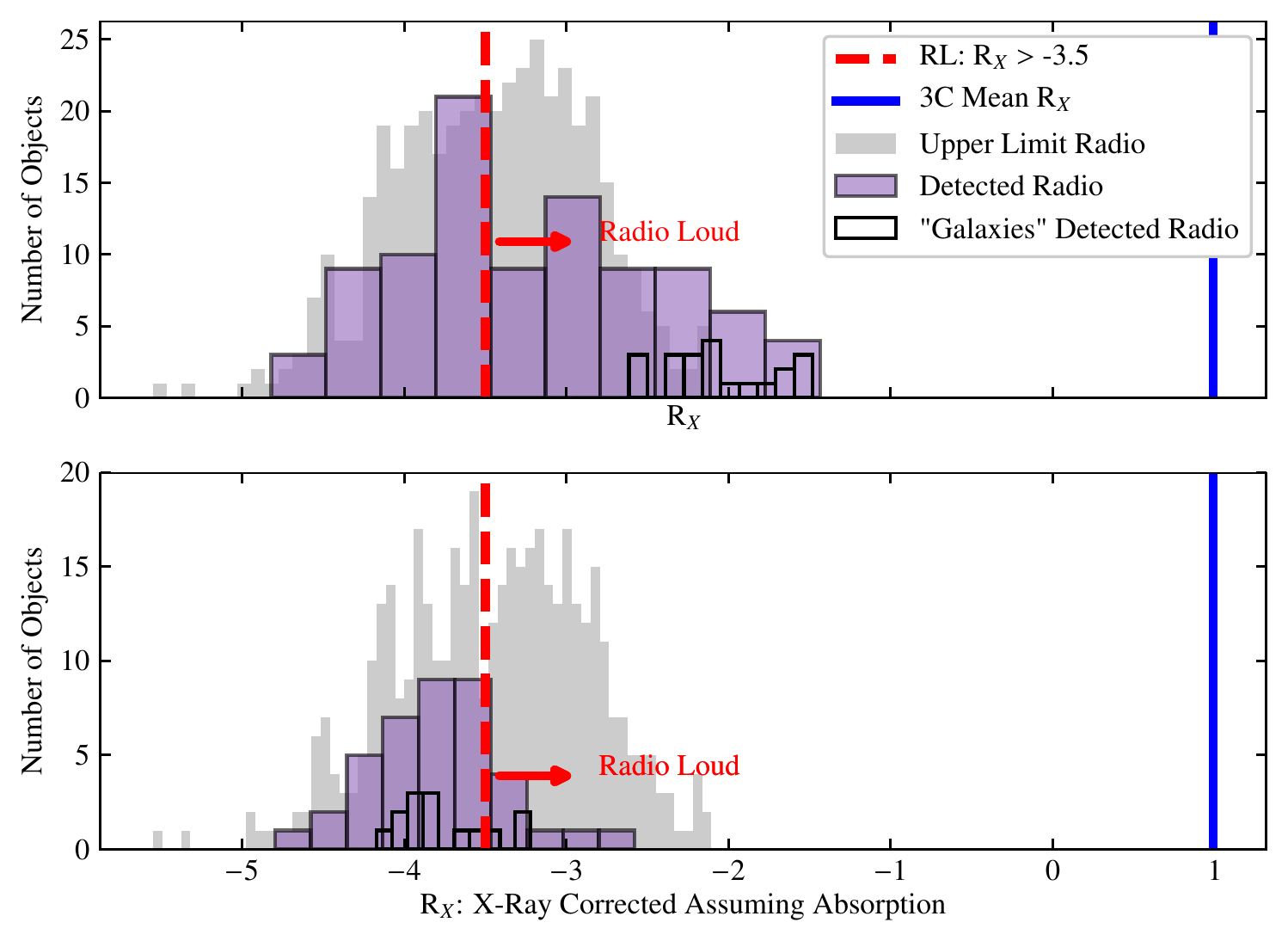}
	\caption{Radio/X-ray radio loudness parameter distribution: In the top panel, we plot the distribution (orange) of the radio-loudness diagnostic, as parametrized by \citet{terashima03}. We calculate the radio upper limits using the limiting flux of the GOODS-S VLA survey \citep{vla} and show the distributions of the upper limit R$_{X}$ (grey). The dashed red line is the RL threshold as empirically found by \citet{terashima03}. For comparison, the blue solid line is the mean R$_{X}$ value for a sample of bonafide RL sources, namely the 3CR sample with $z > 1$ \citep{wilkes13}. In the lower panel, we compute the predicted X-ray luminosities for X-ray under-luminous sources via the L$_{\mathrm{Torus}^{*}}$ values using the S15 relation. The black empty histogram is the distribution for L17 classified "Galaxies" with z $>$ 0.5 and with a detection in the HB.}
\label{fig:rx}
\end{figure*}

\subsection{Estimating the True Obscuration}

We can estimate the level of obscuration by comparing the non-absorption corrected X-ray luminosities to empirical studies utilizing the MIR wavelength measurements. In \autoref{fig:nh_lx}, we determine where the non-absorption corrected luminosities are located within empirically defined regions of non-obscured and heavily obscured sources, indicated by the shaded regions. For the un-obscured region, we use two different intrinsic X-ray - 6 \micron\ AGN luminosity relationships: i) The \citet{gandhi09} relationship, which was derived from a local sample of type 1 AGN, and careful decomposition of the nuclear 6 \micron\ luminosity was performed to minimize host-galaxy contamination ii) The \citet{fiore09} relationship, which was derived from a sample that spanned a larger redshift and X-ray luminosity range as compared to \citet{gandhi09}. The blue shaded region indicates the same relationships but where the X-ray luminosity is absorbed by a column density of $N_\mathrm{H} = 10^{24}$ cm$^{-2}$ as presented in \citet{lansbury15}. For the objects with f$_{X} < 3 \times 10^{-16}$ \fluxunit, $100\%$ are below the empirically shaded region for un-obscured AGN, and 74\% are within or below the $N_\mathrm{H} > 10^{24}$ cm$^{-2}$ parameter space. For these lowest flux objects, 70\% of them have estimated $N_\mathrm{H}$ values that are an order of magnitude greater then the values derived from L17.

We then correct our \lumhardint\ values to account for the underestimation in $N_\mathrm{H}$ by assuming i) L$_{\mathrm{Torus}^{*}}$ is probing predominately AGN processes ii) The lower 2$\sigma$ value of the S15 relationship is a sufficient upper limit of the true intrinsic hard X-ray luminosity. For all objects that are $< 2\sigma$ below the S15 relation in \autoref{fig:lmir_v_lx}, we compute the predicted X-ray luminosity for a given L$_{\mathrm{Torus}^{*}}$ value using the S15 relationship referenced in \autoref{sec:results}. We define these corrected luminosities as \lumprimehardint. 

If our corrected luminosities are a better estimate of the intrinsic luminosity of these AGN, this implies the hardness ratio (HR $=$(FH-FS)/(FH+FS)) for the faintest sources does not provide a correct indication of obscuration. As detailed in \citet{matt1997,matt2000}, a soft scattered component of heavily obscured AGN can dominate at rest energies $<$~10~keV. As we see in Figure 11, our most obscured sources live in the parameter space of moderate to high hardness-ratios. As shown in \citet{brightman2012}, the classical hardness-ratio inference of heavily obscured sources may not be ideal. In L17, the majority of obscured sources do not have enough counts for detailed spectroscopic analysis, and thus, the HR is used to estimate the $N_\mathrm{H}$. In our work, we estimate how much the $N_\mathrm{H}$ would need to be corrected in order to correspond to empirical X-ray-IR relationships. In \autoref{fig:HR_NH}, we combine the hardness ratios, our estimated $N_\mathrm{H}$ (labeled as ``$N_\mathrm{H}$ corrected''), and the X-ray to L$_{\mathrm{Torus}^{*}}$ ratio. We find a fraction of the sources with the highest corrected $N_\mathrm{H}$s, have HRs (0 $-$ 0.5) consistent with X-ray spectra that have a soft-scattered component \citet{brightman2012}.

In the following sections, we explore the effect of these newly derived X-ray luminosities on two important features that are often considered when investigating the nature and the evolution of AGN.

\begin{figure}
\centering
\includegraphics[scale=.6]{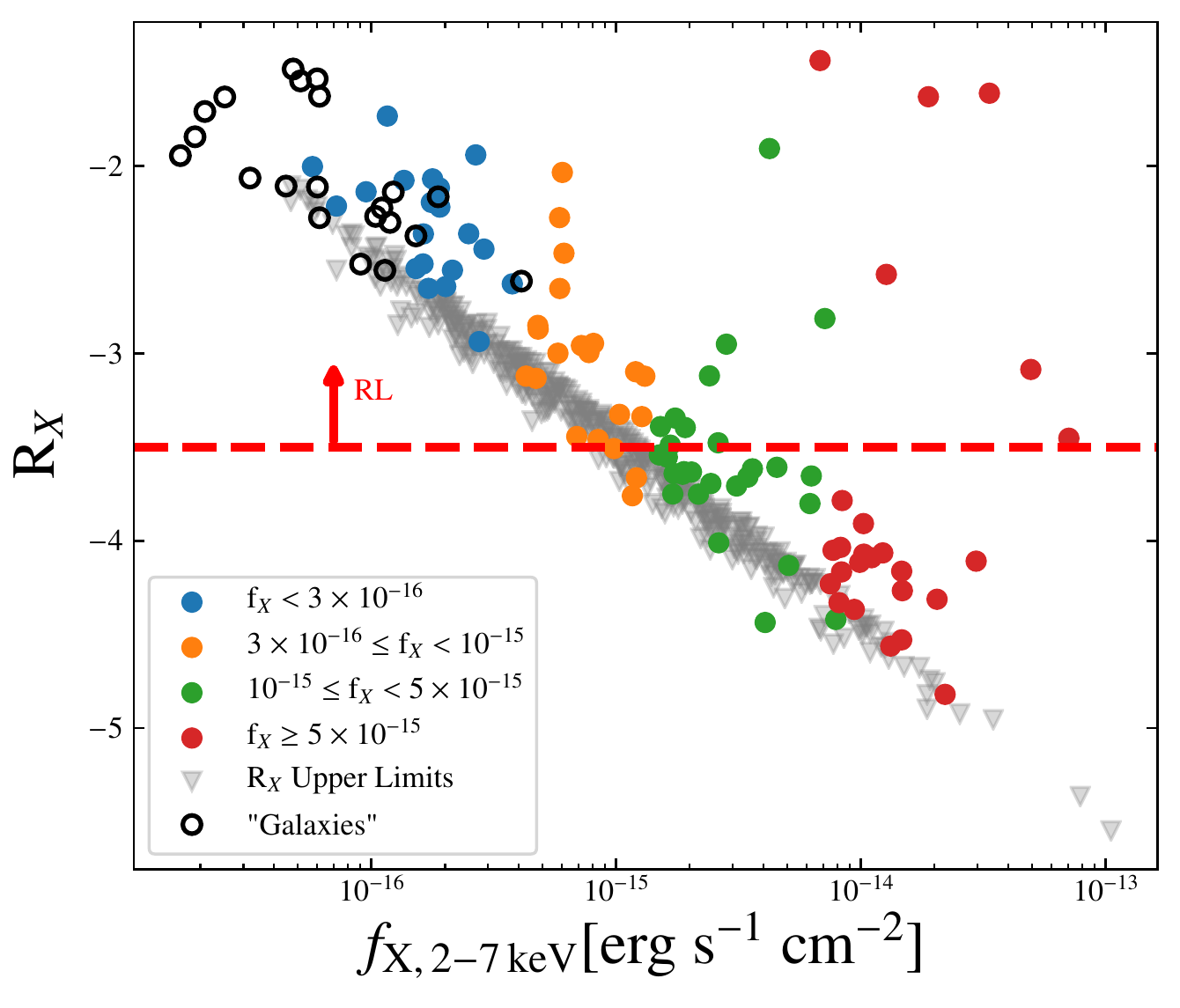}
	\caption{R$_{X}$ as a function of \fluxhard: The red line is the RL threshold as empirically found by \citet{terashima03}, and the upwards arrow indicates the region of radio-loudness. The open black circles are L17 classified "Galaxies" with z $>$ 0.5 and with a detection in the HB.} 
\label{fig:rx_v_flux}
\end{figure}

\subsection{Implications for Radio-Loudness Determination} \label{sec:radio}

About 10\% of AGN have radio emission that is at least 10 times higher than one would expect from SF or other physical processes typical of the majority of AGN \citep{kellerman89}. These objects are known as radio-loud (RL) AGN.  

A wealth of studies have argued for a bi-modality in the distribution of radio-loudness parameters between jetted RL and non-jetted radio-quiet (RQ) AGN \citep{kellerman89,terashima03,padovani2017}. These parameters define radio-loudness as the ratio between the radio luminosity and another luminosity measurement within the spectral energy distribution. With the aim of assessing the presence of a RL population in our sample, we first use the radio-loudness parameter as parametrized by the relationship between the radio luminosity and X-ray luminosity ($R_{X}$ = $\nu L_{\nu,1.4 GHz}$/ L$_{0.5 - 7 keV}$)\citep{terashima03}. This is relevant to this work because sources that are observed as under-luminous in the X-rays with respect to their radio power could be mistakenly identified as RL AGN. If the dimming of  X-ray flux due to the hypothesis of extra obscuration is correct, a large fraction of objects in our sample would be erroneously classified as RL. In fact, a previous analysis of the 4Ms CDFS AGN sample \citep{tozzi}, which included only AGN with L$_{2-10} > 10^{42}$ erg s$^{-1}$, found that roughly 30\% of their objects were RL.

\begin{figure}
\centering
\includegraphics[scale=.6]{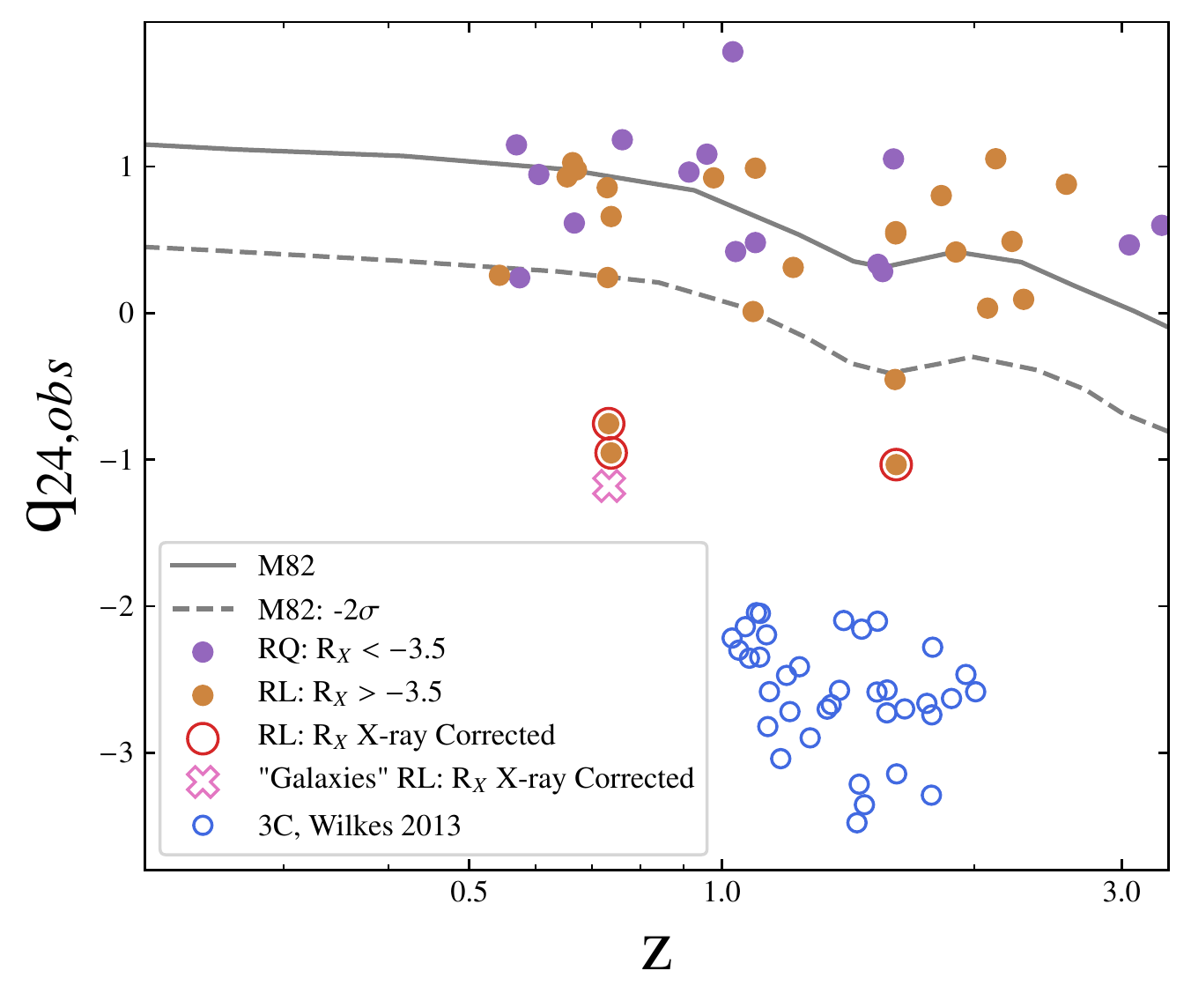}
	\caption{q$_{24,obs}$ as a function of redshift: The orange and blue circle represent the R$_{X}$ RL, R$_{X}$ RQ respectively.The dashed blue line is the lower 2 $\sigma$ evolution of q$_{24,obs}$ for M82 as plotted in \citet{bonzini2012}. The open pink cross is an object within the L17 classified "Galaxies" sub-set. We also plot for reference the high-z 3C objects \citep{wilkes13}.} 
\label{fig:z_v_q24}
\end{figure}

The majority of our L17 sub-sample is not detected in the radio. In \autoref{fig:rx}, we show the distribution of the $R_{X}$ for the 94 sources that are detected at 1.4 GHz \citet{luo17}. For ease of comparison to previous works, we calculate $R_{X}$ using the absorption corrected L$_{0.5 - 7 keV}$ values provided in L17. We calculate the radio luminosities assuming a radio spectral slope of $\alpha = -0.7$ where $f_{\nu} \sim \nu^{\alpha}$. The dashed red-line is the $R_{X}$ threshold for radio-loudness as empirically determined in \citet{terashima03}. The solid blue line is the median R$_{X}$ value for the $z \approx 1$ 3C RL AGN sample \citep{wilkes13}, for reference. The 3C sample is used for comparison because these objects are bona-fide RL AGN with robust X-ray measurements. The grey histogram is the distribution of the upper limit R$_{X}$ for the sources in our sample with a radio upper limit. The radio upper limits are calculated using the limiting flux of the GOODS-S VLA survey \citep{vla}. We also include the the "Galaxies" sub-sample as indicated by the black-edged, transparent histogram. According to the above assumptions, 56\% of the radio detected objects are classified as RL. This is significantly greater than the expected 10\% \citep{terashima03}. Furthermore, in \autoref{fig:rx_v_flux}, we find the majority of objects posing as RL AGN are the sources in the two lowest flux bins. 

Unless the X-ray measurements of our low-flux objects were not significantly underestimated, we would expect a radio-loudness analysis to yield similar number fractions found in other works. As seen in the lower panel of \autoref{fig:rx}, we find the percentage of objects that are classified as RL is significantly reduced when using \lumprimehardint: 13\% out of the 38 objects with radio, MIR, and X-ray detections.  We believe that this constitutes further indication that obscuration is present in a large fraction of these low X-ray flux sources, since this would explain the unreasonable fraction of RL objects observed if obscuration is not correctly taken into account.

\begin{figure*}
\centering
\includegraphics[]{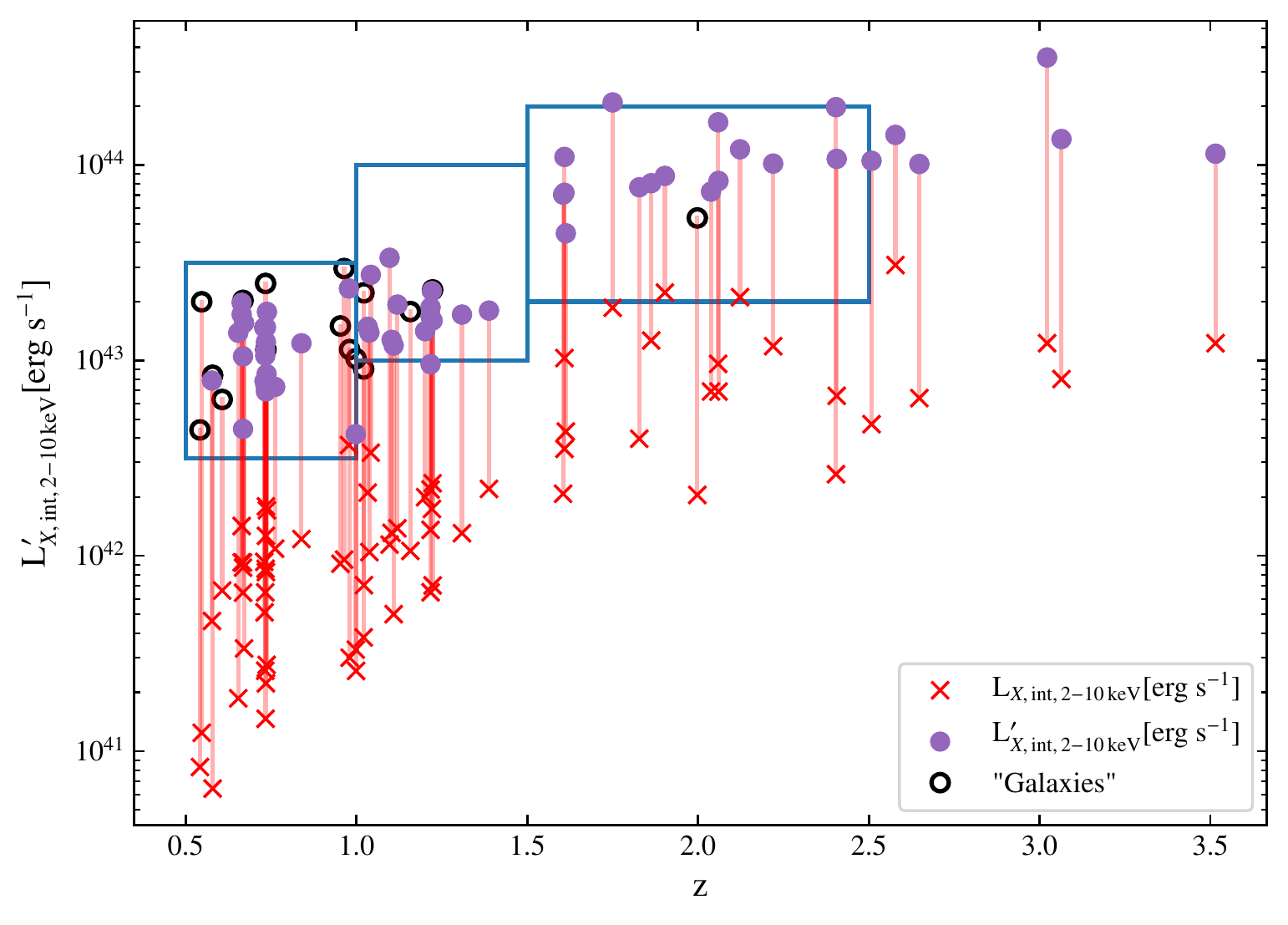}
	\caption{Binned luminosity function analysis: We show the intrinsic hard X-ray as a function of redshift for all sources with the $N_\mathrm{H}$ $>$ 10$^{24}$" cm$^{-2}$ parameter space defined in \autoref{fig:nh_lx}. The blue points are corrected for additional obscuration as defined by \lumprimehardint. The open black circles are L17 classified "Galaxies"}. For comparison, we show the \lumhardint\ values (red crosses). The three rectangles are the three bins used in the luminosity function analysis. 
\label{fig:bindef_lumfun}
\end{figure*}

We can also check radio loudness using a diagnostic that does not rely on the X-ray detections. There are well known correlations between the infrared and the radio \citep{sargent10,bonzini2012} that are expressed through the q$_{24}$ parameter, which is the logarithm of the ratio between the IR and the radio flux density. \citet{bonzini2012} parametrize the q$_{24}$ parameter using the observed 24 \micron\ flux density and observed 1.4 GHz radio flux density:

\begin{equation}
q_{24,obs} = log_{10}(S_{24 \micron}/S_{r})
\end{equation}

where $S_{24 \micron}$ is the observed 24 \micron\ flux density from MIPS and $S_{r}$ is the observed 1.4 GHz flux density from the VLA. 
Observed flux densities are used, rather than rest-frame, due to insufficient data that is needed to derive bolometric values, and to avoid the high uncertainties that are introduced when modeling. \citet{bonzini2012} assume the IR and radio properties of high-redshift star-forming galaxies are similar to local star-forming galaxies. Thus, a template of the prototypical starburst M82 is used to calculate $q_{24,obs}$ as a function of redshift. We use the calculated M82 values as the star-forming galaxy locus via \citet{bonzini2012}, and classify objects that are RL as those with IR to radio fluxes that lie $2 \sigma$ below the SF locus. In \autoref{fig:z_v_q24}, we color-code the q$_{24,obs}$ values by whether they are classified as RL via R$_{X}$. We find significant disagreement between R$_{X}$ and q$_{24,obs}$. Note that only 8\% of our sample is classified as RL when $q_{24,obs}$ is used.

As seen in \autoref{fig:z_v_q24}, the objects that are classified as RL using \lumprimehardint (red circles), there is 100\% overlap with the $q_{24,obs}$ diagnostic. Surprisingly, we also find an object within the “Galaxies” sub-sample that is classified as RL AGN in both the corrected R$_{X}$ diagnostic and q$_{24,obs}$. In summary, without the assumption that there is a significant under-estimation of the X-ray luminosity, over half of our sample would be erroneously classified as RL.

\subsection{Implications for Obscured AGN Space Density}

The results discussed in this paper also have important bearings for cosmological studies. A major implication of our finding is in fact related to the space density of obscured AGN. We estimate the space density for the obscured sources in our sample with and without the corrected X-ray luminosities. We use a binned luminosity function, and define the three bins as: $10^{42.5} <$ \lumgen\ $\leq 10^{43.5}$, $10^{43} <$ \lumgen\ $\leq 10^{44}$, $10^{43.3} <$ \lumgen\ $\leq 10^{44.3}$ and redshifts $0.5 < z \leq 1.0$, $1.0 < z \leq 1.5$, $1.5 < z \leq 2.5$ respectively, and where all luminosity units are in erg s$^{-1}$. In \autoref{fig:bindef_lumfun}, we show the intrinsic hard X-ray luminosity as a function of redshift for all sources with $N_\mathrm{H}> 10^{24}$~cm$^{-2}$ as defined by being within or below the blue-shaded region in \autoref{fig:nh_lx}. The blue points are corrected for additional obscuration as defined by \lumprimehardint. The open black circles are L17 classified "Galaxies". For comparison, we show the \lumhardint values (red crosses). We choose the luminosity-redshift bins to maximize the number of  sources included in the calculation, while minimizing the number of  potential outliers. We immediately find that the difference between the \lumhardint\ and \lumprimehardint\ values would have a significant effect on space density calculations. Furthermore, we can quantify this effect by comparing the space density of our most obscured sources to model expectations.

We take all of our objects with an estimated $N_\mathrm{H} > 10^{24}$~cm$^{-2}$, and calculate the space density of our heavily obscured sources in the CDFS field. We present two space densities per luminosity, redshift bin. The first is the AGN sub-sample presented in \autoref{fig:lmir_v_lx}, and the second includes these sources plus the objects in "Galaxies" sub-sample. We use a binned luminosity function as parametrized by \citet{ranalli16}. The differential luminosity function $\Phi$ is defined the number of objects $N$ at co-moving volume $V$:
\begin{equation}
    \Phi(L,z) = \frac{d^{2}N(L,z)}{dVdL}
\end{equation}

We approximate the LF within a bin with luminosity boundaries L$_\mathrm{min, 2-10 keV}$, L$_\mathrm{max, 2-10 keV}$ and redshift boundaries $z_\mathrm{min}$, $z_\mathrm{max}$ as $N/V_\mathrm{probed}$ where $V_\mathrm{probed}$ is:
\begin{equation}
    V_{probed} = \int_{L_{min}}^{L_{max}} \int_{z_{min}}^{z_{max}} \Omega(L,z) \frac{dV}{dz} dz dL,
\end{equation}
$dV$/$dz$ is the co-moving volume, and $\Omega(L,z)$ is the survey coverage at the flux that an object of luminosity $L$ would have if placed at redshift $z$. 

\begin{figure}
\centering
\includegraphics[scale=.8]{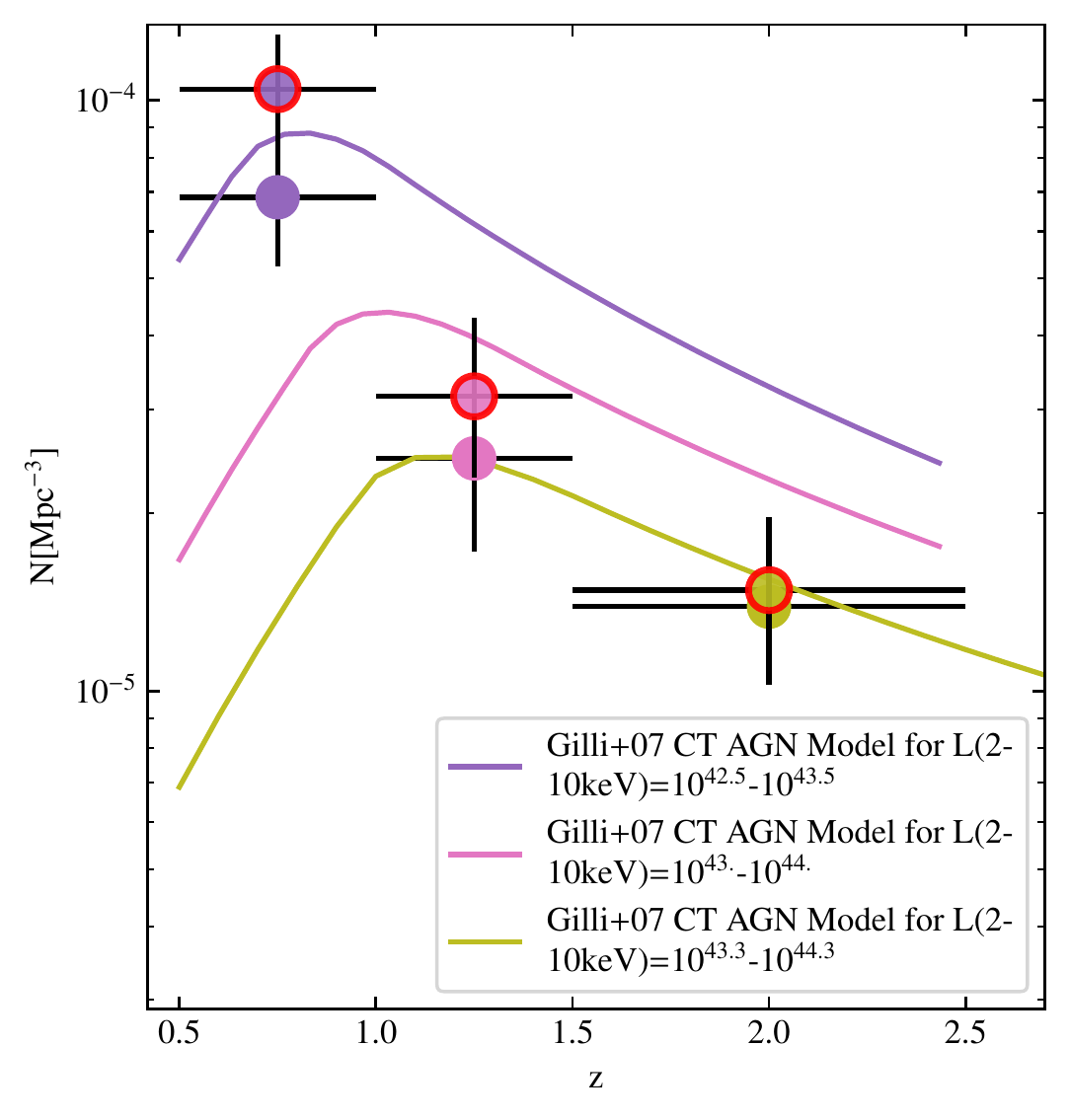}
	\caption{AGN Space Density: The solid colored three points are the heavily obscured AGN space densities for each luminosity-redshift bin, and their colors correspond to the theoretical model values for that bin. The points with red-circles are space density estimates including the L17 classified "Galaxies" sub-sample. The errors include model error and the upper and lower ends of the 68.3\% confidence interval estimated using the Gehrels approximation \citep{gehrels}. The solid lines are the expected functions for three luminosity bins as modelled by \citet{gilli07}.} 
\label{fig:agndens}
\end{figure}

In \autoref{fig:agndens}, we show the space densities as a function of redshift for the three luminosity-redshift bins. The AGN only sub-sample contains 19, 11, and 15 heavily obscured AGN candidates in the three redshift bins respectively. The estimated space density of these objects within the redshift and luminosity bins are be $6.9 \pm 1.5 \times 10^{-5}$ Mpc$^{-3}$, $2.5 \pm 0.9 \times 10^{-5}$ Mpc$^{-3}$, and $1.4 \pm 0.6 \times 10^{-5}$ Mpc$^{-3}$. The AGN + "Galaxies" combined sub-sample contains 29, 14, and 16 objects for the three redshift bins. The estimated space density of the "Galaxies" combined sub-sample within the redshift and luminosity bins are $12.8 \pm 3.2 \times 10^{-5}$ Mpc$^{-3}$, $4.2 \pm 0.9 \times 10^{-5}$ Mpc$^{-3}$, and $1.5 \pm 0.6 \times 10^{-5}$ Mpc$^{-3}$. The errors include the upper and lower ends of the 68.3\% confidence interval estimated using the standard Gehrels approximation \citep{gehrels}. We note that the $x$-axis errors in \autoref{fig:agndens} represent the range of the redshift bin used in the space density calculation. We find agreement with the predicted space density functions calculated using the X-ray background in \citet{gilli07}. The models were based on the X-ray luminosity function observed at low redshift and parametrized with a luminosity dependent density evolution. The \citet{gilli07} models were computed between the redshift and luminosity intervals referenced in \autoref{fig:agndens}, and between $10^{24}$ cm$^{-2} < $ N$_{\mathrm{H}}$ $< 10^{26}$ cm$^{-2}$. CT AGN that do not contribute to the X-ray background probed in \citet{gilli07}, such as sources with low or zero scattering fractions or sources with an obscuring medium that have a $4\pi$ covering factor, would not be represented, and thus these models represent lower limits. 

The sources used to calculate the space densities in \autoref{fig:agndens} were measured to have $N_\mathrm{H}> 10^{24}$~cm$^{-2}$, we estimate the  error on this assumption by comparing the difference in space density estimates when using objects only below the below shaded region to the values derived in \autoref{fig:agndens}. We find a maximum 15\% difference between including all of the objects in the blue-shaded region versus only the objects below the blue-shaded region.  

We also estimate the number of AGN that may be missed in 7Ms CDFS via comparing to GOODS-S Spitzer/Herschel IR maps in this region \citet{elbaz11}.  As seen in \citet{donley12}, only 52\% of IRAC AGN have X-ray counterparts in the 50-150ks \textit{Chandra} exposures. We check if there are a significant portion of IRAC AGN lacking Chandra 7MS counterparts, for these may represent the most obscured AGN in the GOODS-South field. We first choose a sub-field of the IRAC and Chandra GOODS-South images, where both maps overlap with one another. We then identify the IRAC AGN sources using \citet{donley12} power-law AGN criterion: x $>$ 0.08  and y$>$0.15  and  y $>$(1.12 $\times$ x)-0.27  and  y$<$(1.12 $\times$ x)+0.27  and  f4.5 $>$ f3.6  and f5.8 $>$ f4.5 and f8.0 $>$ f5.8 where x=log(f5.8/f3.6) and y=log(f8.0/f4.5). We find 48 power-law AGN in this IRAC GOODS-S sub-field. We then one to one match these sources to the Chandra 7MS cut-out using their GOODS-S IDs. We find 38 power-law AGN that also have a Chandra 7MS soft, full, and/or hard band detection. Thus, we find 80\% of IRAC AGN have an X-ray counterpart when compared to the significantly deeper 7Ms exposure. While this shows we may be missing 10 objects in this sub-field due to lack of an X-ray detection, it also shows the amount of total AGN in the IR missed by the 7Ms CDFS survey is within $\sim$20\%. We are aware that this is still an incomplete assessment, for a more accurate estimate should take into account the different selection biases between IR and X-ray catalogues. However, we remind the reader that the main goal of this paper is to improve upon characterizing the obscured AGN population for X-ray selected sources.

Furthermore, the lowest and highest redshift bin space density enables us for the first time to make an accurate comparison with models in a parameter space poorly explored thus far. If we did not consider objects from the lowest X-ray flux bins as being obscured AGN, the estimated space density in the lowest redshift bin would drop by 50\% and the highest redshift bin would drop by 40\%. By taking into account the results of our work, we are able to probe a fainter luminosity bin then previously estimated in the literature, and we find both heavily obscured AGN space density calculations consistent with the X-ray background models.

\section{Summary and Conclusions} \label{sec:conclusion}

Utilizing the excellent wavelength coverage of the GOODS-South field, we compare the X-ray luminosities of AGN from the \textit{Chandra} 7Ms survey to the radio (VLA 1.4 GHz), optical grism spectroscopy (\textit{HST}-WFC3), high resolution optical/NIR imaging and photometry (\textit{HST}-ACS, \textit{HST}-WFC3IR), and NIR/MIR/FIR photometry (\textit{Spitzer} IRAC,\textit{Spitzer} IRS PUI, \textit{Spitzer} MIPS, \textit{Herschel} PACS). We find the lowest X-ray flux AGN (f$_{X}$ $<$ 3 $\times$ 10$^{-16}$ \fluxunit) in our sample have the greatest disagreement with their X-ray luminosities compared to their radio, infrared, and optical counterparts. 

We estimate the AGN contribution to the MIR by redshift correcting the observed IRAC 8~\micron, IRAC PUI 16~\micron\ and MIPS 24~\micron\ fluxes for objects whose redshift corresponds to luminosities in the range between 3.2~\micron\ to 5.7~\micron. Of these objects, 44\% are at least 2$\sigma$ below the expected S15 relationship which defines the relationship for absorption corrected AGN in the MIR and X-ray. 

The interpretation of these low-flux sources with under-estimated X-ray luminosity, is that a large column of obscuring material ($N_\mathrm{H}>10^{23}$~cm$^{-2}$) is attenuating the X-ray emission. Assuming these objects are indeed obscured AGN, we find that almost all of the lowest X-ray flux AGN in our L17 sub-sample have $N_\mathrm{H} > 10^{24}$~cm$^{-2}$. 

We explore the implications of our results, and choose two examples where under-estimated X-ray luminosities could affect AGN research. Using the radio diagnostics of \citet{terashima03} and \citet{bonzini2012}, 56\% of our objects have \lumhardint\ that would place in the radio-loud regime as compared to their 1.4~GHz radio emission. When we correct our X-ray luminosities for additional obscuration only 13\% of our objects are classified as RL. For the sources with an estimated $N_\mathrm{H}>10^{24}$ cm$^{-2}$ we calculate the heavily obscured AGN space density in the following luminosity-redshift bins: $10^{42.5} <$ \lumgen\ $\leq 10^{43.5}$, $10^{43} <$ \lumgen\ $\leq 10^{44}$, $10^{43.3} <$ \lumgen\ $\leq 10^{44.3}$ and redshifts $0.5 < z \leq 1.0$, $1.0 < z \leq 1.5$, $1.5 < z \leq 2.5$ respectively. We find the heavily obscured AGN space densities for these bins to be $6.9 \pm 1.5 \times 10^{-5}$ Mpc$^{-3}$, $2.5 \pm 0.8 \times 10^{-5}$ Mpc$^{-3}$, and $1.4 \pm 0.4 \times 10^{-5}$ Mpc$^{-3}$. Our results are in agreement with models of the obscured AGN space density function as derived by \citet{gilli07}. 

Future work to test our estimation of the level of intrinsic obscuration can occur with not only future missions, but also with currently operating telescopes. Using a large ground based telescope, we can obtain more sensitive [\textsc{Oiii}] measurements, as well as other optical emission lines to further probe the AGN power. In addition, we can use ALMA to characterise the dustiness of the host galaxies. This would test whether the un-accounted for obscuration is truly located within parsecs of the SMBH versus host galaxy obscuration (see \citealt{circosta19,damato2020} for further examples of this possibility). Future X-ray missions, such as \textit{ATHENA}, will enable more sensitive X-ray measurements. This would allow for more rigorous spectral analysis of the low X-ray flux sources. Finally, \textit{JWST} will allow us to directly image the MIR flux on kpc scales. Thus, we could more robustly decouple SF from torus emission.  

In conclusion, we find a significant fraction of the low flux population of \textit{Chandra} 7Ms AGN are obscured AGN in disguise. This population is usually missed and/or mis-classified and should be taken into account when constructing AGN samples from deep X-ray surveys.

\section*{Acknowledgements}
We thank the anonymous referee for their thoughtful insight and important contributions to this work. In addition, we thank Julian Krolik, Alexandra Pope, Arianna Brown, Duncan Watts, Kirsten Hall, and Raymond Simons for useful discussions and insight. ELL is supported by the Maryland Space Grant Consortium. RG acknowledges support from the agreement ASI-INAF n. 2017-14-H.O. This research has made use of the NASA/IPAC Infrared Science Archive, which is operated by the Jet Propulsion Laboratory, California Institute of Technology, under contract with the National Aeronautics and Space Administration. This publication makes use of data products from the \textit{Wide-field Infrared Survey Explorer}, which is a joint project of the University of California, Los Angeles, and the Jet Propulsion Laboratory/California Institute of Technology, funded by the National Aeronautics and Space Administration. This publication makes use of data products from the Two Micron All Sky Survey, which is a joint project of the University of Massachusetts and the Infrared Processing and Analysis Center/California Institute of Technology, funded by the National Aeronautics and Space Administration and the National Science Foundation. We acknowledge the extensive use of the following Python packages: \software{pandas, scipy, ipython, matplotlib} \citep[respectively]{pandas, scipy,ipython,matplotlib}. This research made use of \texttt{astropy}, a community-developed core Python package for Astronomy \citep{astropy}.

\clearpage
\bibliography{references}

\end{document}